\documentclass[twocolumn,amsmath]{aastex702}

\newcommand{\uv}{$uvby\text{--}\beta$\,}

\graphicspath{{./}{figures/}}

\begin{document}
	
	\title{\uv\ photoelectric photometry of FGK-type pulsating star candidates in the \textit{Kepler} field: observations, stellar parameters and  asteroseismic analysis}
	
	\author[orcid=0000-0002-2781-873X, sname="Fox-Machado", gname="Lester"]{L. Fox-Machado}
	\affiliation{Universidad Nacional Autónoma de México, Instituto de Astronomía, AP 106, Ensenada 22800, BC, México}
	\email[show]{lfox@astro.unam.mx}  
	\begin{abstract}
		
		We present \uv
		photoelectric photometry of 218 FGK-type stars brighter
		than $Kp = 12$ mag in the \textit{Kepler} field.
		Interstellar reddening was estimated using the observed
		photometric indices, distances from \textit{Gaia} DR3,
		and the Bayestar19 three-dimensional dust maps.
		The sample was classified into dwarf and giant stars,
		and their metallicities and effective temperatures
		were determined using empirical photometric calibrations.
		The resulting parameters were compared
		with those reported in previous photometric and
		spectroscopic studies.		
		We also analyzed the \textit{Kepler} long-cadence
		light curves of the sample and determined the global
		asteroseismic parameters, including the frequency
		of maximum oscillation power ($\nu_{\max}$) and the
		large frequency separation ($\Delta\nu$), for red
		giants exhibiting solar-like oscillations.		
		Finally, we performed a Bayesian isochrone-fitting
		analysis using PARSEC stellar evolutionary models,
		incorporating the photometric metallicities and
		effective temperatures, the asteroseismic parameters,
		and absolute visual magnitudes derived from
		\textit{Gaia} DR3 data as observational constraints.
		This analysis provides estimates of the fundamental
		stellar parameters, including effective temperatures,
		metallicities, surface gravities, masses, and radii,
		together with their associated uncertainties.
		
	\end{abstract}


\section{INTRODUCTION}
\label{sec:intro}

During its primary mission, the \textit{Kepler} space telescope monitored a fixed 105 deg$^2$ field of view (FOV) between the Cygnus and Lyra constellations. The resulting long-term, high-precision light curves for diverse stellar types constitute an invaluable database that continues to be actively exploited, particularly for probing stellar interiors by means of asteroseismology. Since determining the properties of target stars within the \textit{Kepler} FOV is essential for asteroseismic studies, numerous ground-based follow-up campaigns have been conducted using both spectroscopic \citep[e.g.,][]{uytt2011, decat2015, pinso2014, pinso2018, pinso2025, ren2016, zong2018} and photometric \citep[e.g.,][]{casagrande2014} techniques. Such initiatives aim to refine fundamental stellar parameters and cross-validate the values reported in the \textit{Kepler} Input Catalog (KIC). Contributing to this goal, we focused on characterizing stars in the \textit{Kepler} FOV by means of Str\"omgren-Crawford photometry.

The \uv photometric system \citep{stromgren1963, stromgren1964, crawford1970, crawford1975} comprises four medium-width filters ($uvby$) alongside the $\beta$ index to determine stellar atmospheric parameters. The key indices are defined as $(b-y)$,  $m_1 =  (v-b) - (b-y)$, and $c_1 = (u-v) - (v-b)$. The colour $(b-y)$  is relatively insensitive to blanketing effects and thus is well suited for measuring effective temperatures when reddening is known \citep[e.g.,][]{alonso1996, alonso1999, clem2004, ramirez2005, casagrande2010}. The $m_1$ index quantifies metal-line blanketing near 4100 \AA, serving as a metallicity ($[\mathrm{Fe/H}]$) indicator for a
variety of stars \citep[e.g.,][]{stromgren1964,olsen1984, schuster1989, ramirez2005, casagrande2010, casagrande2011,hilker2000, ramirez2004, calamida2007, calamida2009, casagrande2014}. The $c_1$ index is designed to evaluate the Balmer discontinuity, which is a temperature indicator for B- and A-type stars and  a surface gravity indicator for later type stars, and a tracer of evolutionary stages when combined with other $uvby$ indices \citep[e.g.,][]{crawford1975, olsen1984, nissen1991, schuster2004, twarog2007}.
Additionally, the extinction-free $\beta = \beta_{w} - \beta_{n}$ index measures  the strength of the hydrogen $H_\beta$ line, serving as a key indicator of temperature and luminosity.
Here, we present new \uv photoelectric photometry for a number of FGK-type pulsating candidate stars brighter than $K_{\rm p}=12$ mag,  derive  physical parameters and investigate the evolutionary states of them by combining ground-based data with space observations and evolutionary models.

\begin{figure}
	\begin{center}
		\includegraphics[scale=0.35]{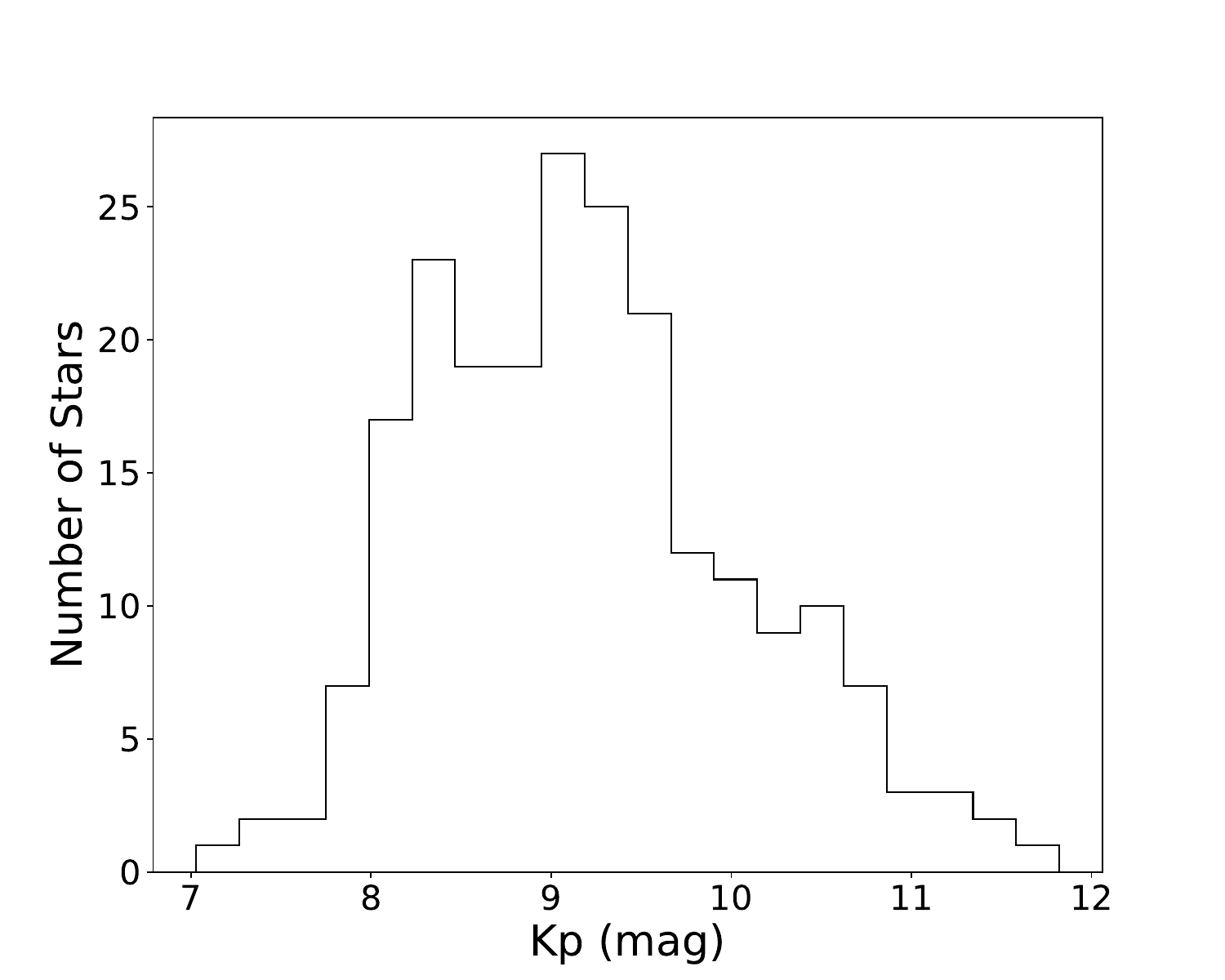}
	\end{center}
	\caption{$Kp$ magnitude distribution of the sample.}
	\label{fig:histog}
\end{figure}

\section{The Stellar Sample}
\label{sec:sample}

Our targets were selected based on their apparent brightness and spectral types. As illustrated in Fig.~\ref{fig:histog}, most of the targets have \textit{Kepler} magnitudes in the range $Kp \sim 8$--$11$ mag. The brightest stars are often saturated in CCD observations, even with the shortest available exposure times on small telescopes. Consequently, although these stars are particularly suitable for precise asteroseismic modeling and can benefit from multiple independent observational constraints, they are often excluded from ground-based CCD Str\"omgren photometric surveys \citep[e.g.,][]{casagrande2014}.

\section{Observations and Data Reduction}
\label{sec:observations}

The \uv data were acquired using the 0.84m telescope at the Observatorio Astronómico Nacional en San Pedro Mártir (OAN-SPM), Mexico, during two observing runs: 2013 (July 27 to August 04, 9 nights) and 2014 (August 04 to 09, 6 nights). Because our observational and reduction methodologies have been extensively documented in previous works \citep[e.g.,][]{schuster1988, schuster1989, schuster2004, schuster2016, pena2006, pena2007, pena2014, pena2017, pena2023, fox2025}, we summarize them only briefly here. 

We employed a six-channel photoelectric photometer. This instrument enables simultaneous observations in the four $uvby$ channels or, when switched to its $\beta$ mode, simultaneous measurements in the narrow and wide $\beta$ channels. The $uvby$ section operates as a spectrograph-photometer, utilizing exit slots and optical interference filters to define the band-passes. To ensure precise spectral alignment on the exit slots within approximately $\pm 1\,\mathrm{\mathring{A}}$, the grating angle was calibrated on the first night of each run using a cadmium lamp.  To determine atmospheric extinction coefficients and time-dependent night corrections, extinction stars were observed nightly over an airmass range of at least 0.8. Additionally, several "drift" stars were observed symmetrically with respect to the local meridian (two hours east and two hours west) throughout each night whenever possible.  Each measurement comprised at least six 10-second integrations per star and three 10-second integrations of the sky for the $uvby$ and $H_\beta$ filters. To maintain high signal-to-noise ratios, program stars were observed to at least 50,000 counts in all four $u,v,b,y$ channels and at least to 30,000 counts in $\beta$ channels. Sky background was measured until its statistical contribution became negligible.  These count levels provide photon-noise uncertainties of only a few millimagnitudes per channel, consistent with an overall photometric precision of approximately 0.01 mag.

Data reduction followed the numerical procedures described by \citet{arellano1988}. During each night a number of standard stars, selected primarily from \citet{olsen1983, olsen1984, olsen1993} and \citet{hauk1998} with a large span in magnitude and color indices were observed in order to establish a nightly instrumental photometric system considering atmospheric extinction corrections and linear time-dependent terms. The instrumental measurements were then transformed to the standard $uvby\text{--}\beta$ system using the linear transformation equations of \citet{crawford1966} and \citet{crawford1970} and  the color terms in the calibration equations were derived. In particular,  linear $(b-y)$ terms were included in the $m_1$ and $c_1$ transformation equations to account for bandwidth effects in the $V$ filter. Finally, the instrumental $y$ measurements were calibrated to the standard Johnson $V$ magnitude system \citep{johnson1953}.

We assessed the calibration quality for each observing season by analyzing the residuals, defined as the differences between the published standard values and the transformed instrumental measurements of the standard stars. During the 2013 observing season, 38 standard stars were observed in a total of 302 individual measurements. The mean instrumental photometric uncertainties ranged from 0.004 mag in $(b-y)$ to 0.007 mag in $c_1$, while the RMS residuals of the transformation equations ranged from 0.005 to 0.018 mag, respectively. During the 2014 observing season, 37 standard stars were observed in 187 individual measurements. The corresponding mean instrumental uncertainties ranged from 0.005 mag in $(b-y)$ to 0.014 mag in $c_1$, with RMS transformation residuals between 0.005 and 0.021 mag. In both observing seasons, the uncertainties in the $V$ magnitude, $m_1$ index, and $\beta$ index were intermediate between the ranges reported for $(b-y)$ and $c_1$.
On the other hand, 34 standard stars utilized in  2013 were also observed in 2014. The mean absolute differences in magnitudes and colours ($2013 - 2014$) for $V$, $(b-y)$, $m_1$  and $c_1$ are 0.003, 0.001, 0.001 and 0.002 mag, respectively.  Overall, these results demonstrate the stability and accuracy of the photometric transformations, which has also been extensively validated in previous works.

\section{Str\"omgren-Crawford indices of the sample}
\label{sec:indices}

The full set of Str\"omgren--Crawford photometric indices, $V$, $(b-y)$,
$m_1$, $c_1$, and $\beta$, for the target stars is presented in
Table~\ref{tab:uvby_data}, together with their corresponding uncertainties. For stars with two or more independent measurements, the quoted uncertainties correspond to the standard deviation of the individual measurements. This applies to 89 stars in the $uvby$ photometry and to 152 stars in the H$\beta$ measurements. For stars with only one photometric measurement, the internal uncertainty was  estimated from the dispersion of the six individual 10-s integrations obtained in each filter. The standard error of the mean count rate in each filter was first computed and converted into a magnitude uncertainty. These individual uncertainties were then propagated in quadrature through the definitions of the Str\"omgren indices to obtain $\sigma_{V,{\rm inst}}$, $\sigma_{(b-y),{\rm inst}}$, $\sigma_{m_1,{\rm inst}}$, $\sigma_{c_1,{\rm inst}}$. For H$\beta$, the uncertainties in the narrow- and wide-band measurements were similarly derived from the six integrations and propagated in quadrature to obtain $\sigma_{\beta, \rm inst}$. Finally, for single-measurement stars, the uncertainty associated with the
transformation to the standard system for the corresponding night was added
in quadrature to the internal instrumental uncertainty: $\sigma_{\rm ind} =
\left(\sigma_{\rm inst}^{2}+\sigma_{\rm std}^{2}\right)^{1/2}$. Thus, the uncertainties listed in Table~\ref{tab:uvby_data} represent the
observed scatter for stars with repeated measurements, whereas for
single-measurement stars they represent the propagated internal precision of
the integrations combined with the uncertainty of the nightly standard-system transformation.

{For stars with multiple independent measurements, we compared the standard deviations of the repeated observations with the uncertainties estimated for a single measurement, which include the internal instrumental and transformation errors. The median differences (standard deviation of repeated measurements minus the estimated uncertainty for a single measurement), calculated individually for each star, are $-0.021$, $-0.0035$, $-0.0018$, $-0.0032$, and $-0.0064$ mag for $V$, $(b-y)$, $m_1$, $c_1$, and H$\beta$, respectively. Similarly, the median differences between the standard deviations of repeated measurements and the uncertainties associated with the transformations to the standard system are $-0.0013$, $+0.0020$, $+0.0023$, and $-0.0042$ mag for $V$, $(b-y)$, $m_1$, and $c_1$, respectively. These comparisons indicate that the different uncertainty estimates are generally of the same order of magnitude, with the combined uncertainties typically exceeding the observed scatter between independent measurements.}

 The Str\"omgren--Crawford indices and  \textit{Kepler} light curves  were used as complementary
diagnostics to distinguish giant stars from dwarfs. On this
basis, the final sample was divided into 95 giants and 123 dwarfs.

\begin{table*}[!t]
	\caption{Photometric $uvby\text{--}\beta$ data for the target stars.}
	\label{tab:uvby_data}
	\centering 
	\setlength{\tabcolsep}{5.5pt}
	
		\begin{tabular}{rcccccccccccc}
			\hline
			\hline
			KIC ID & $V$ & $\sigma_V$ & $(b-y)$ & $\sigma_{b-y}$ & $m_1$ & $\sigma_{m_1}$ & $c_1$ & $\sigma_{c_1}$ & $N_{uvby}$ & $\beta$ & $\sigma_{\beta}$ & $N_\beta$ \\
			\hline
			757137 & 8.332 & 0.032 & 0.459 & 0.007 & 0.163 & 0.009 & 0.429 & 0.012 & 1 & 2.492 & 0.056 & 1 \\
			1430163 & 9.610 & 0.032 & 0.282 & 0.007 & 0.148 & 0.008 & 0.495 & 0.012 & 1 & 2.647 & 0.001 & 2 \\
			1435467 & 8.976 & 0.032 & 0.340 & 0.007 & 0.167 & 0.008 & 0.504 & 0.012 & 1 & 2.630 & 0.021 & 2 \\
			2161731 & 8.445 & 0.032 & 0.601 & 0.007 & 0.478 & 0.007 & 0.309 & 0.011 & 1 & 2.584 & 0.004 & 2 \\
			2305930 & 11.131 & 0.032 & 0.680 & 0.009 & 0.329 & 0.011 & 0.382 & 0.017 & 1 & 2.616 & 0.001 & 2 \\
			2309550 & 9.521 & 0.032 & 0.640 & 0.007 & 0.471 & 0.008 & 0.397 & 0.013 & 1 & 2.593 & 0.009 & 2 \\
			2833697 & 8.994 & 0.032 & 0.591 & 0.008 & 0.261 & 0.010 & 0.367 & 0.015 & 1 & 2.567 & 0.016 & 1 \\
			2837475 & 8.509 & 0.032 & 0.282 & 0.007 & 0.155 & 0.007 & 0.517 & 0.011 & 1 & 2.683 & 0.013 & 2 \\
			2969462 & 8.443 & 0.032 & 0.644 & 0.007 & 0.430 & 0.007 & 0.321 & 0.011 & 1 & 2.596 & 0.002 & 2 \\
			3123191 & 9.888 & 0.032 & 0.330 & 0.007 & 0.144 & 0.007 & 0.444 & 0.011 & 1 & 2.628 & 0.019 & 2 \\
            \hline
\end{tabular}	
\tablecomments{This table is published in its entirety in machine-readable format. A portion is shown here for guidance regarding its form and content.}
\end{table*}

\subsection{Comparison with Other Str\"omgren studies}
\label{sec:comp_ind}

A CCD-based Str\"omgren photometric study of the \textit{Kepler} field
was presented by \citet{casagrande2014}. Using the Wide Field Camera
on the 2.5-m Isaac Newton Telescope (INT) at La Palma, they observed a
strip centered at a Galactic longitude of $74^{\circ}$ and covering
latitudes from approximately $8^{\circ}$ to $20^{\circ}$, and reported $(b-y)$, $m_1$, and $c_1$ indices for 989 red giants exhibiting solar-like oscillations. We identified 22 giant stars in common between their catalog and our sample. For these stars, the mean differences (this work minus \citealt{casagrande2014}) in $(b-y)$, $m_1$, and $c_1$ are $-0.003$, $-0.001$, and $-0.010$ mag, with standard deviations of 0.03, 0.05, and 0.04 mag, respectively. The small mean offsets, all below 0.01 mag in absolute value, indicate good agreement between the photometric zero points of the two studies, although the star-to-star scatter reaches 0.03--0.05 mag.

\section{Reddening and intrinsic photometric quantities}
\label{sec:reddening}

The reddening across the \textit{Kepler} field was investigated in
detail by \citet{casagrande2014}. They determined distances by combining
asteroseismic radii with angular diameters obtained through the infrared
flux method (IRFM), and estimated $E(B-V)$ using the three-dimensional
extinction models of \citet{amores2005} and \citet{drimmel2003}. Their
reddening scale was empirically calibrated using the open cluster
NGC~6819. We note that $H\beta$ photometry was not available for their
analysis.

In the present work, accurate \textit{Gaia}~DR3 distances are available
for the entire sample. We therefore combined our \uv\ photometry
with the Bayestar19 three-dimensional dust map \citep{green2019}.
The reddening measures reported by the map were converted to 
Johnson $E(B-V)_{\rm Bayestar19}$ values by applying a conversion factor of 0.884, based on the Pan-STARRS1 $E(g-r)_{\rm P1}$ color and an $R_V=3.1$
reddening law \citep{green2019}.

Inspection of the resulting Bayestar19 reddenings showed that many of
the nearby stars, particularly at $d\lesssim 300$ pc, are assigned
zero reddening. In contrast, the Str\"omgren \uv\
photometry yields small but non-zero color excesses for many of these
objects. We therefore used a subset of stars with reliable photometric
reddening determinations as empirical reference stars to derive a local
correction to the Bayestar19 reddenings.

\begin{figure}
	\begin{center}
		\includegraphics[scale=0.35]
		{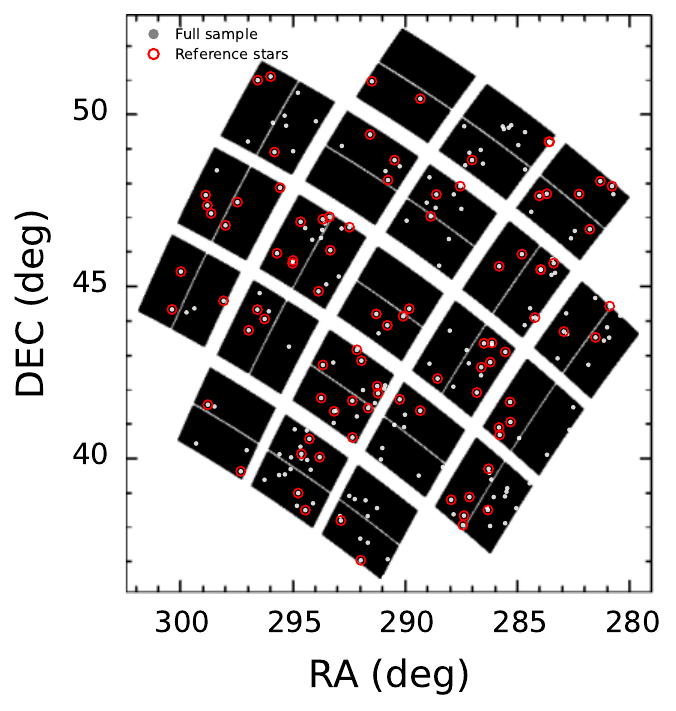}
	\end{center}
	\caption{Spatial distribution of the observed stellar sample across
		the \textit{Kepler} field of view. Small circles indicate the
		full sample, while open circles identify the 92 reference stars
		with reddening determinations obtained from the \citet{moon1985}
		\texttt{uvbybeta} calibration. These stars are distributed
		throughout the \textit{Kepler} focal plane and provide broad angular
		coverage for deriving local empirical corrections to the Bayestar19
		reddenings.}
	\label{fig:kepler_fov}
\end{figure}

Individual $E(b-y)$ values were derived from the empirical \uv\
calibrations implemented in the \citet{moon1985}\footnote{The calibration of Moon (1985) was adopted because it provides a more practical approach for our analysis. The alternative calibration of \citet{schuster1989} was developed primarily for F- and G-type main-sequence stars, whereas our sample includes both dwarfs and giants.} \texttt{uvbybeta}
code and converted to the Johnson system using the standard relation $
	E(B-V)={E(b-y)}/{0.74}$. 	Although a complete set of Str\"omgren--Crawford indices is available for our sample, the applicability of the
	\citet{moon1985} calibration depends on both spectral type and
	luminosity class. The \citet{moon1985} routines use $(b-y)$, $m_1$,
	$c_1$, and $\beta$, together with a group number between 1 and 8 that
	selects the appropriate region of the H--R diagram for the star.	
	In this classification scheme, Group 7 applies to F1--G2 stars of
	luminosity classes III--V, whereas Group 8 extends to later spectral
	types, G2--M2, but only for luminosity classes IV--V. Consequently,
	giants later than G2 are not covered by the adopted Moon classification
	scheme and no direct Str\"omgren reddening was derived for them.	
	We found 92 stars in the F1--G2 spectral type that satisfy the criteria for 
	Group 7 of the Moon classification scheme.  This group includes
	luminosity classes III--V and is defined by
	$2.60 < \beta < 2.72$ and
	$0.22 < (b-y)_0 < 0.39$.  These stars have reliable Str\"omgren reddening determinations,
	and were therefore adopted as the reference sample for deriving the
	local corrections to the Bayestar19 reddenings. 
	 The distances of the reference stars range from approximately 41 to 474 pc, with a median distance of 180 pc; 82 of them are located within 300 pc. Their distribution across the \textit{Kepler} field is
shown in Figure~\ref{fig:kepler_fov}.

For each reference star, we calculated the residual between the
Str\"omgren and Bayestar19 reddening estimates,

\begin{equation}
	\Delta E(B-V) =
	E(B-V)_{\rm Str\ddot{o}mgren}
	-
	E(B-V)_{\rm Bayestar19}.
\end{equation}

\noindent The reference sample has a mean residual of
$\Delta E(B-V)=+0.0231$ mag and a median of $+0.0212$ mag, with a
standard deviation of 0.0197 mag. The positive offset is mainly
associated with nearby stars for which Bayestar19 returns very small
or zero reddening. Because the reference stars are distributed across the full
\textit{Kepler} field of view (Figure~\ref{fig:kepler_fov}), their
residuals provide empirical information on local differences between
the Str\"omgren and Bayestar19 reddening estimates over a wide range
of lines of sight. The positions of the reference stars were
transformed from Galactic longitude, latitude, and \textit{Gaia}
distance into three-dimensional Galactic Cartesian coordinates
$(X,Y,Z)$. For each target without a direct Str\"omgren reddening
determination, the five nearest reference stars in this
three-dimensional space were identified. A local correction was then
estimated from the inverse-distance-weighted mean of their individual
Str\"omgren--Bayestar19 residuals.

For each neighboring reference star, the weight was defined as
$w_i = \frac{1}{d_i^2}$, where $d_i$ is the three-dimensional spatial separation between the target and the $i$th reference star. The inverse-square
dependence is used solely as a weighting prescription, assigning
greater weight to reference stars located closer to the target; it
does not imply a physical $1/d^2$ dependence of interstellar
reddening.

The local Str\"omgren--Bayestar19 correction assigned to each target
was therefore calculated as

\begin{equation}
	\Delta E(B-V)_{\rm local} =
	\frac{\displaystyle\sum_{i=1}^{5}
		w_i\,\Delta E(B-V)_i}
	{\displaystyle\sum_{i=1}^{5}w_i},
\end{equation}

\noindent where

\begin{equation}
	\Delta E(B-V)_i =
	E(B-V)_{{\rm Str\ddot{o}mgren},i}
	-
	E(B-V)_{{\rm Bayestar19},i}.
\end{equation}

The corrected reddening for each target was then obtained as

\begin{equation}
	E(B-V)_{\rm corrected} =
	E(B-V)_{\rm Bayestar19}
	+
	\Delta E(B-V)_{\rm local}.
\end{equation}

\noindent This procedure assumes that the
Str\"omgren--Bayestar19 residual varies smoothly over the local spatial
scales sampled by the reference stars, so that spatially neighboring
stars are expected to have similar corrections. The inverse-square
distance weighting causes the nearest reference stars to dominate the
local estimate while progressively reducing the contribution from
more distant neighbors.

The reliability of this procedure was evaluated directly using the
reference stars themselves. Each reference star was considered in
turn as a target without a Str\"omgren reddening determination and was
temporarily excluded from the reference set. Its local
$\Delta E(B-V)$ correction was then calculated from the five nearest
stars among the remaining 91 references, using exactly the same
inverse-square-distance weighting described above. This correction was
added to the Bayestar19 reddening of the star to obtain an independent
$E(B-V)_{\rm corrected}$. The procedure was repeated for all 92
reference stars, allowing the corrected reddening to be compared with
the independently determined Str\"omgren reddening for every star. The resulting differences, $\delta E(B-V) = E(B-V)_{\rm Str\ddot{o}mgren} - E(B-V)_{\rm  corrected}$, have a mean of $+0.0003$ mag, an rms dispersion of
0.0209 mag, and a median absolute deviation of 0.0135 mag.
Therefore, within the volume sampled by the reference stars, the
local correction to the Bayestar19 maps equates the empirical Str\"omgren reddening scale, without introducing a significant systematic offset. 

For the final adopted reddenings, the directly determined Str\"omgren
values were retained for the reference stars. For the remaining
stars within the distance range sampled by the reference sample
($d\leq474$ pc), the Bayestar19 reddening was adjusted using the
local inverse-distance-weighted correction defined above. Beyond
474 pc,  the original $E(B-V)_{\rm Bayestar19}$  reddening was adopted.

The uncertainties in the adopted reddenings were estimated according to
the source of the reddening determination. 
The uncertainty in the Bayestar19 reddening, denoted by \(\sigma_{\rm Bayestar19}\), was estimated from the posterior samples provided by the dust map as half the difference between the 84th and 16th percentiles. For stars within the distance range covered by the Strömgren reference sample, the uncertainty in the corrected reddening was calculated by combining \(\sigma_{\rm Bayestar19}\) in quadrature with the uncertainty of the local empirical correction. The latter accounts for both the weighted spatial dispersion of the Strömgren–Bayestar19 residuals among the five nearest reference stars and the propagated uncertainties of their reddening measurements. For the Strömgren reference stars themselves, 
a representative uncertainty
of $\sigma_{E(b-y)}=0.020$ mag was adopted.  This is a conservative choice
compared with the average uncertainty of
$\sigma_{E(B-V)}\simeq0.020$ ($\sigma_{E(b-y)}\simeq0.015$)mag estimated by
\citet{casagrande2014} from the scatter of their photometric reddening
determination.  For stars beyond the maximum distance of the reference sample, no local correction was applied; consequently, the final reddening uncertainty was simply \(\sigma_{\rm Bayestar19}\).

\begin{figure}
	\begin{center}
		\includegraphics[scale=0.35]
		{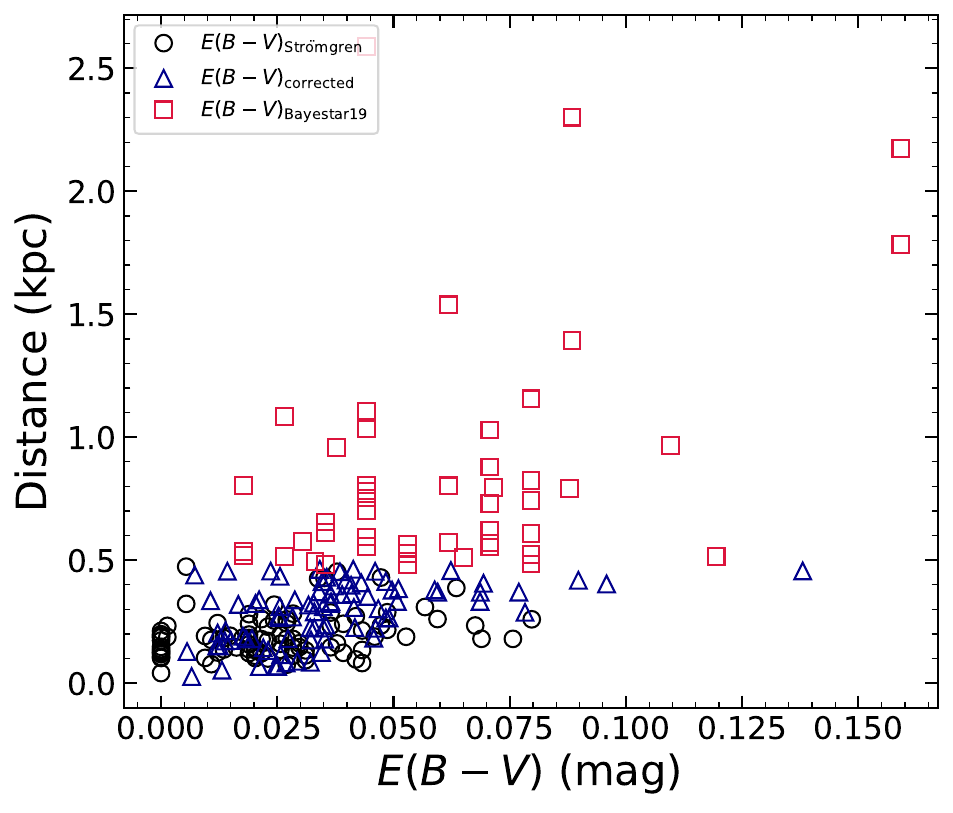}
	\end{center}
	\caption{\textit{Gaia}~DR3 distance as a function of the adopted
		reddening $E(B-V)$. Circles represent the 92 reference stars
		with Str\"omgren reddenings, triangles show the locally
		corrected Bayestar19 reddenings for stars within the reference-star
		distance range, and squares indicate the Bayestar19
		reddenings adopted for stars beyond 474 pc.}
	\label{fig:ebv_dist_gaia}
\end{figure}

We note that at larger distances, KIC~9655101, a confirmed
member of the open cluster NGC~6811, has a Bayestar19 reddening of
$E(B-V)=0.044$ mag at a \textit{Gaia} distance of 1.11 kpc. This value
is consistent, within the uncertainties, with the independent reddening
determination for NGC~6811 of $E(B-V)=0.066 \pm 0.025$ mag \citep{janes2013}, providing an independent check on the adoption of  unmodified Bayestar19 values beyond the volume sampled by the reference stars. Figure~\ref{fig:ebv_dist_gaia} shows the final adopted reddenings as a function of \textit{Gaia}~DR3 distance for the complete sample.

\begin{figure}
	\begin{center}
		\includegraphics[scale=0.35]
		{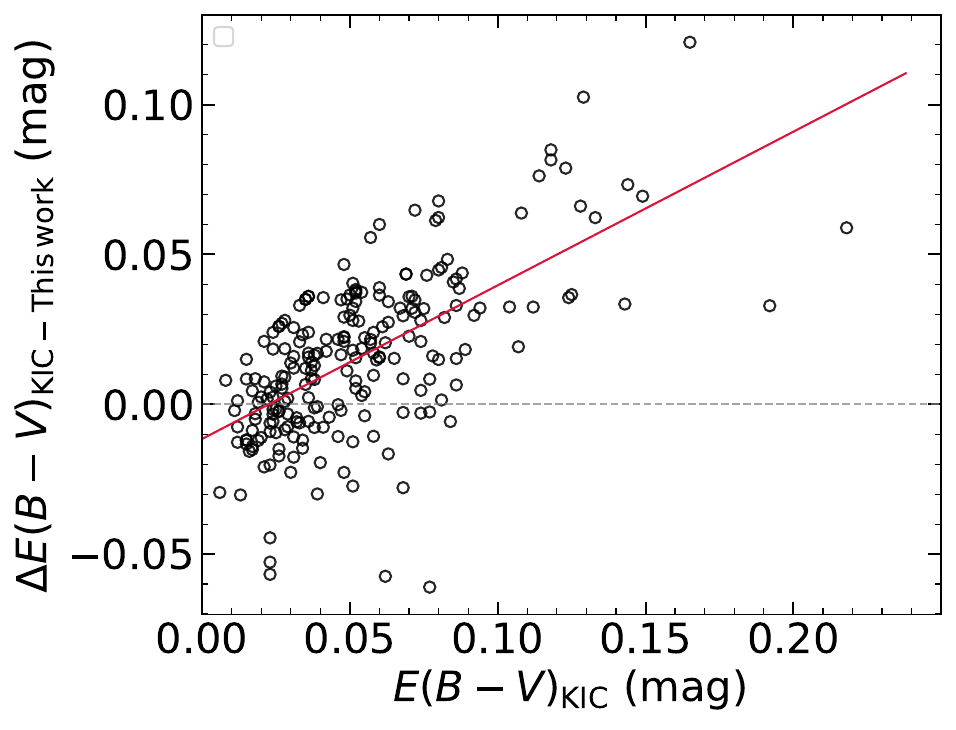}
	\end{center}
	\caption{Comparison between the color excesses $E(B-V)$ reported
		in the KIC and those adopted in this work. The continuous line represents a linear fit to observational data.}
	\label{fig:comp_ebv}
\end{figure}

In Figure~\ref{fig:comp_ebv}, we compare our adopted $E(B-V)$ values
with the color excesses reported in the KIC. A systematic trend is
apparent, with the KIC increasingly overestimating the reddening toward
larger $E(B-V)$. A similar behavior was found by
\citet{casagrande2014}, who showed that the discrepancy between their
SAGA reddenings and those reported in the KIC increases with increasing
KIC color excess.

We dereddened the Str\"omgren photometry using the derived reddenings
and the standard interstellar extinction relations from
\citet{crawford1976}:

\begin{align}
	V_0 &= V - 4.28 \, E(b-y), \\
	(b-y)_0 &= (b-y) - E(b-y), \\
	m_0 &= m_1 + 0.32 \, E(b-y), \\
	c_0 &= c_1 - 0.20 \, E(b-y),
\end{align}

\noindent where $V_0$ is the dereddened visual magnitude, and
$(b-y)_0$, $m_0$, and $c_0$ are the intrinsic Str\"omgren indices. The uncertainties in the dereddened quantities were obtained by propagating
in quadrature the observational errors and the
uncertainties in the adopted $E(b-y)$ values derived in the previous section.

\begin{figure}
	\begin{center}
		\includegraphics[scale=0.35]{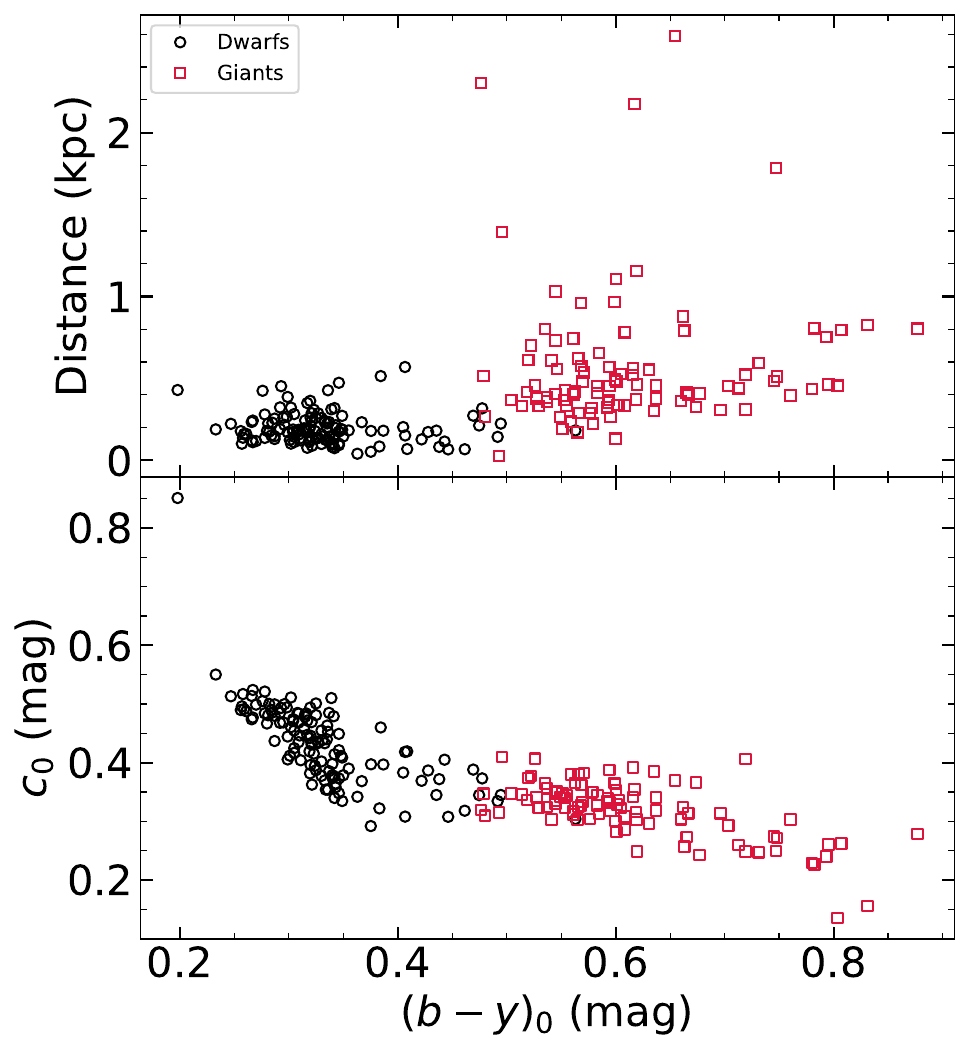}
	\end{center}
	\caption{Top: \textit{Gaia} DR3 distance as a function of the
		intrinsic $(b-y)_0$ color for the stellar sample. Bottom: $c_0$ as a function of
		$(b-y)_0$. Circles and squares represent dwarfs and giants,
		respectively; {the two luminosity groups show different distributions, although some overlap is present.} The uncertainties in the \textit{Gaia} DR3 distances are relatively small that they are omitted for clarity. }
	\label{fig:v_dist_gaia}
\end{figure}

Figure~\ref{fig:v_dist_gaia} shows the distribution of the sample in
the $(b-y)_0$ versus \textit{Gaia} distance plane (top panel) and in
the $c_0$ versus $(b-y)_0$ diagram (bottom panel). Circles and squares
represent dwarfs and giants, respectively. {The two luminosity groups show
different distributions, although some overlap is present.}

\bigskip

\section{Metallicity and effective temperature from empirical calibrations}
\label{sec:calibrations}

\bigskip

The determination of stellar metallicity from photometry has evolved significantly 
since the pioneering work of Str\"omgren. In early formulations \citep{stromgren1964,crawford1975,olsen1984}, stellar metallicity $[\mathrm{Fe/H}]$ was estimated using the differential parameter $\delta m_1 = m_{1,\mathrm{std}} - m_{1,\mathrm{obs}}$, which quantifies how much stellar metal-line blanketing departs from a fiducial baseline at a given temperature. This reference sequence was traditionally defined by Hyades cluster members in the $m_1$--$(b-y)$ plane, allowing chemical abundance variations to be isolated from effective temperature effects. Similarly, the pressure-sensitive index $c_1$ was evaluated via $\delta c_1 = c_{1,\mathrm{std}} - c_{1,\mathrm{obs}}$ relative to the zero-age main sequence (ZAMS) to account for surface gravity variations.

 Modern empirical calibrations relate $[\mathrm{Fe/H}]$ directly to the intrinsic indices $(b-y)_0$, $m_0$, and $c_0$,  typically
  calibrated against high-resolution spectroscopic iron abundances. Such relations have been established for both dwarf populations \citep{olsen1984,schuster1989,haywood2002,nordstrom2004,twarog2007,casagrande2011} and  giant stars \citep{hilker2000,faria2007,calamida2007,aden2011,casagrande2014}.

The derived $[\mathrm{Fe/H}]$ values can subsequently be incorporated into color--temperature relations
which are  parameterized in terms of the reciprocal effective temperature, $\theta_{\rm eff} \equiv 5040/T_{\rm eff}$, using a polynomial expansion in intrinsic color and $[\mathrm{Fe/H}]$:

\begin{equation}
	\theta_{\rm eff} =
	a_0 + a_1 X + a_2 X^2 +
	a_3 X[\mathrm{Fe/H}] +
	a_4[\mathrm{Fe/H}] +
	a_5[\mathrm{Fe/H}]^2,
\end{equation}

\noindent where $X$ represents the color index, and the $a_i$ are coefficients determined from empirical calibration samples \citep[e.g.,][]{ramirez2005}. 

The calibrations adopted in the present work are listed in Table~\ref{tab:calib}. Metallicities were obtained from the relations of \citet{casagrande2011} for dwarfs and \citet{casagrande2014} for giants, while effective temperatures were determined using \citet{casagrande2010} and \citet{ramirez2005}, respectively. The color--temperature relations are based on effective-temperature scales established through the infrared flux method (IRFM), which has a relatively weak dependence on stellar atmosphere models \citep[e.g.,][]{ramirez2005,casagrande2010}. 

The color indices used in Equation~(9) for each luminosity class are listed in the last column of Table~\ref{tab:calib}. For giant stars, we adopted the near-infrared color $(V-K_s)_0$ instead of $(b-y)_0$, as the latter is more sensitive to metal-line blanketing in the atmospheres of cool stars. The dereddened $(V-K_s)_0$ colors were calculated by combining our measured $V$ magnitudes with  $K_s$ photometry from the Two Micron All Sky Survey \citep[2MASS;][]{skrutskie2006}, applying the corresponding interstellar extinction corrections.

The uncertainties in $[\mathrm{Fe/H}]$ and $T_{\rm eff}$ include both the errors propagated from the relevant photometric quantities (see Table~\ref{tab:calib}) and the intrinsic scatter of the corresponding empirical calibration, added in quadrature. 

\begin{table*}[!ht]
	\caption{Empirical calibrations applied to the different luminosity-class groups.}
	\label{tab:calib}
	\begin{center}
		\begin{tabular}{lccr}
			\hline
			Parameter & Sample & Calibration used in this work & Photometric quantities \\
			\hline
			$[\mathrm{Fe/H}]$ & Dwarfs & \citet{casagrande2011} & $(b-y)_0,m_0,c_0$ \\
			$[\mathrm{Fe/H}]$ & Giants & \citet{casagrande2014} & $(b-y)_0,m_0,c_0$ \\
			$T_{\rm eff}$ & Dwarfs& \citet{casagrande2010} & $(b-y)_0,[\mathrm{Fe/H}]$ \\
			$T_{\rm eff}$ & Giants & \citet{ramirez2005} & $(V-K_s)_0,[\mathrm{Fe/H}]$ \\
			\hline
		\end{tabular}
	\end{center}
\end{table*}

\subsection{Comparison with other works}

\begin{figure*}
	\begin{center}
		\includegraphics[scale=0.35]
		{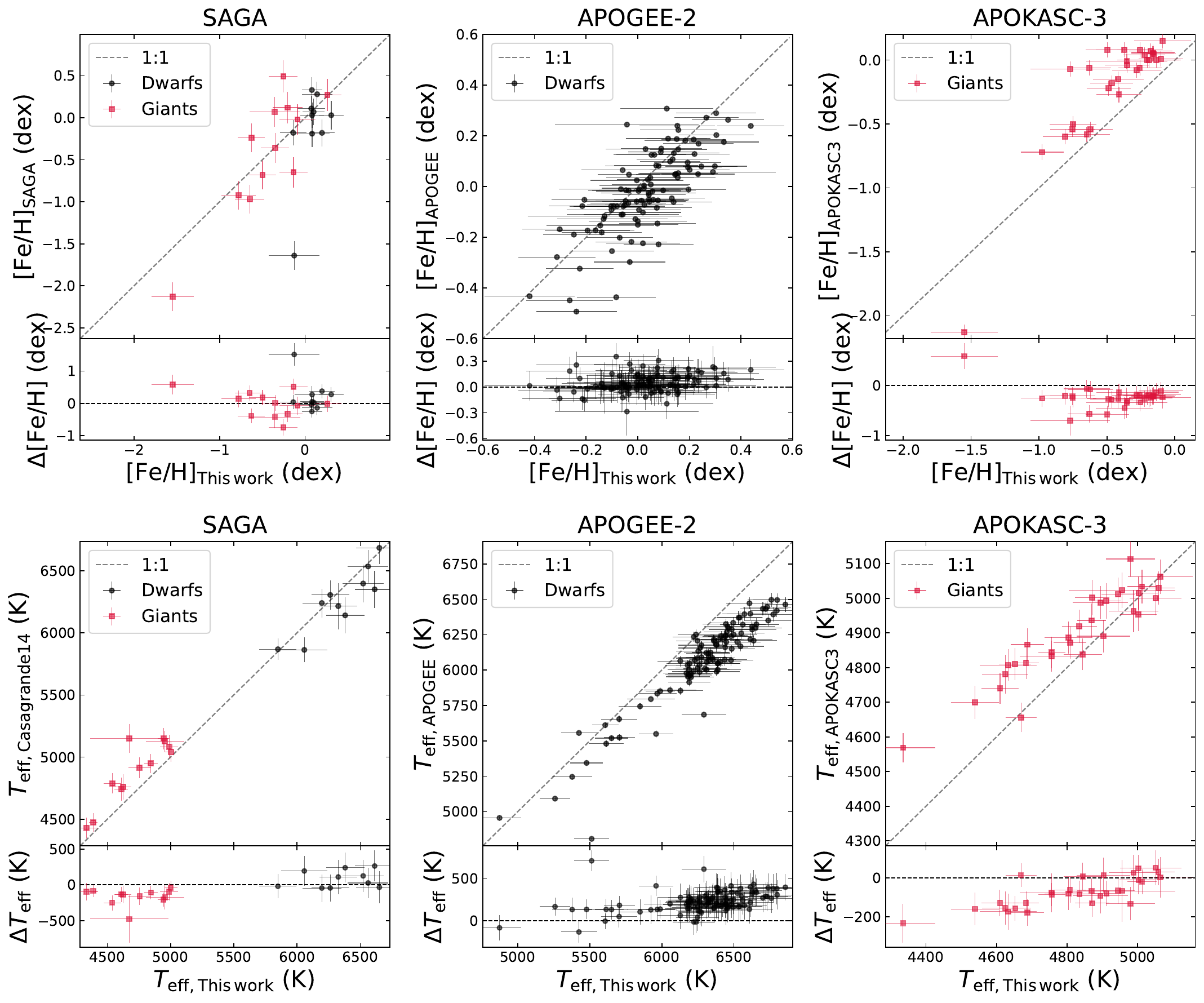}
	\end{center}
	\caption{Comparison of the photometric metallicities (upper panels) and effective temperatures (lower panels) derived in this work with values from SAGA (left column), APOGEE-2 (center column), and APOKASC-3 (right column) for the stars in common with each sample. The residual panels show the differences defined as $\Delta = {\rm This\ work} - {\rm reference}$.}
	\label{fig:comp_feh_teff}
\end{figure*}

Figure~\ref{fig:comp_feh_teff} compares our photometric metallicities and effective temperatures with those from SAGA \citep{casagrande2014} (dwarfs and giants), APOGEE-2 \citep{abdurrouf2022} (dwarfs), and APOKASC-3 \citep{pinso2025} (giants).

For the 22 stars in common with SAGA, the mean metallicity difference (this work minus SAGA) is $+0.09$ dex, with a standard deviation of $0.45$ dex and a median difference of $+0.03$ dex. Two outliers, visible in the upper-left panel of Fig.~\ref{fig:comp_feh_teff}, contribute substantially to the large dispersion. Excluding these objects, the mean difference decreases to $-0.01$ dex, with a standard deviation of $0.31$ dex and a median of $+0.02$ dex. The dwarf outlier has several independent spectroscopic metallicity determinations in the literature, most of which indicate a slightly subsolar metallicity. The giant outlier is also the most metal-poor star in our comparison with APOKASC-3. For this star, we derive $[\mathrm{Fe/H}]=-1.55\pm0.25$ dex, compared with $-2.13\pm0.06$ dex reported by APOKASC-3 (see below).

Apart from these two outliers, our photometric metallicities show reasonable agreement with the SAGA determinations, with the measurements distributed around the one-to-one relation and no appreciable mean offset. 
This agreement is consistent with the small mean differences in the Str\"omgren indices found for the same stars in Sect.~\ref{sec:comp_ind}.

For the same sample, the mean effective temperature difference (this work minus SAGA) is $\Delta T_{\rm eff}=-51\pm169$ K, with a median of $+64$ K (lower-left panel of Fig.~\ref{fig:comp_feh_teff}). When dwarfs and giants are considered separately, the mean differences are $\Delta T_{\rm eff}=+82\pm121$ K (median $+69$ K) and $\Delta T_{\rm eff}=-162\pm114$ K (median $-133$ K), respectively, where the quoted dispersions represent the standard deviations of the differences. These results indicate that, within the limited common sample, our effective temperatures for dwarfs are in reasonable agreement with the SAGA values, whereas those for giants tend to be systematically lower. The two stars showing the largest metallicity discrepancies do not exhibit comparably large differences in $T_{\rm eff}$.

\medskip

A more independent assessment is provided by comparisons with spectroscopic catalogs. For the 104 dwarfs in common with APOGEE-2, the mean metallicity difference (this work minus APOGEE-2) is $+0.05\pm0.11$ dex (median $+0.04$ dex), indicating good agreement with the spectroscopic determinations (upper-middle panel of Fig.~\ref{fig:comp_feh_teff}). In contrast, our effective temperatures are systematically higher than the APOGEE-2 values, with a mean difference of $\Delta T_{\rm eff}=+228\pm123$ K (median $+215$ K; lower-middle panel of Fig.~\ref{fig:comp_feh_teff}). The quoted dispersions represent the standard deviations of the differences. The temperature offset is larger than that found relative to SAGA, although the latter comparison is based on a considerably smaller sample.

For the 30 giants in common with APOKASC-3, our metallicities are systematically lower than the spectroscopic values, with a mean difference of $-0.24\pm0.21$ dex (median $-0.23$ dex; upper-right panel of Fig.~\ref{fig:comp_feh_teff}). In their comparison with APOGEE, \citet{casagrande2014} found that the SAGA--APOGEE metallicity differences depended on $[\mathrm{Fe/H}]$, rather than exhibiting a nearly constant offset. Their comparison, however, was based on an earlier APOGEE data release and a larger sample. Given the limited number of giants in our comparison, the offset observed here may represent only a restricted portion of the metallicity-dependent trend reported by \citet{casagrande2014}, although differences in sample selection and spectroscopic data releases must also be considered.

The effective temperatures of the same giants show reasonably good agreement with APOKASC-3, with a mean difference of $\Delta T_{\rm eff}=-73\pm76$ K (median $-79$ K; lower-right panel of Fig.~\ref{fig:comp_feh_teff}), indicating that APOKASC-3 temperatures are slightly higher on average. \citet{casagrande2014} also reported good overall agreement between the SAGA and APOGEE temperatures, with $\Delta T_{\rm eff}=+90\pm105$ K, but with the opposite sign, as APOGEE yielded lower temperatures than SAGA. The agreement was particularly good below approximately 4600 K, whereas the differences increased toward higher temperatures. It should be noted, however, that the SAGA--APOGEE and our APOKASC-3 comparisons span different effective temperature ranges, approximately 3800--5300 K and 4300--6800 K, respectively, with an overlap of about 1000 K.

{The systematic differences in effective temperature may be partly attributed to the different temperature scales adopted. In particular, the calibration of \citet{ramirez2005}, used for our giant stars, yields temperatures approximately 100 K cooler than the IRFM-based scale of \citet{casagrande2010}, adopted in the SAGA survey \citep{casagrande2014}. This difference could partially explain the lower temperatures obtained for our giants. For the dwarf stars, the larger differences with APOGEE-2 may be related to the temperature calibrations adopted in that survey, which incorporate near-infrared photometric information, as well as to differences in the adopted reddening corrections. These factors could contribute to the observed systematic offsets between the temperature scales.}

\section{Global asteroseismic parameters for red giant stars}
\label{sec:seismic}

Solar-like oscillations are standing sound waves that are stochastically
excited and intrinsically damped by turbulent convection in the outer
layers of stars. They are observed in stars with convective envelopes,
including solar-type main-sequence stars, subgiants, and red giants
\citep{aerts2010}. Their oscillation spectra exhibit a characteristic
frequency pattern that can be described, to first order, by two global
asteroseismic parameters: the frequency of maximum oscillation power,
$\nu_{\rm max}$, and the large frequency separation, $\Delta\nu$.
The former approximately scales with surface gravity and effective
temperature, whereas the latter scales with the square root of the mean
stellar density \citep{bedding2003}.

{ We downloaded the available {\it Kepler} light curves of our sample stars from the Mikulski Archive for Space Telescopes (MAST),
including both long- and short-cadence observations when available. From the FITS files, we extracted the PDCSAP fluxes,
corrected for instrumental systematics by the {\it Kepler} Pre-search Data Conditioning pipeline, together with their corresponding time stamps.
After removing invalid measurements, the fluxes were normalized to the median value of each observing segment and combined into a
single time series for each target. The relative flux was then expressed in ppm, and the power spectral density (PSD) was computed using the
Python {\rm Lightkurve} package. The light curves and their corresponding power spectra were visually inspected. A reliable oscillation power excess was identified for 79 red giants, for which the subsequent asteroseismic analysis was performed.}

The observed PSD of each star was represented by a model consisting of
three Harvey-like components with a fixed exponent of four
\citep{harvey1985}, a white-noise contribution, and a Gaussian
oscillation envelope \citep[e.g.,][]{kallinger2010},

\begin{equation}
P(\nu) =
\sum_{i=1}^{3}
\frac{A_i}{1+(\nu/B_i)^4}
+
H_{\rm env}
\exp\left[
-\frac{(\nu-\nu_{\rm max})^2}
{2\sigma_{\rm env}^2}
\right]
+ W ,
\end{equation}

\noindent where $A_i$ and $B_i$ characterize the amplitude and
characteristic frequency of each Harvey-like component, respectively,
and $W$ represents the white-noise level. The parameters
$H_{\rm env}$ and $\sigma_{\rm env}$ are the height and width of the
oscillation envelope, while its central frequency provides an estimate
of $\nu_{\rm max}$.

\begin{figure}
	\begin{center}
		\includegraphics[scale=0.35]
		{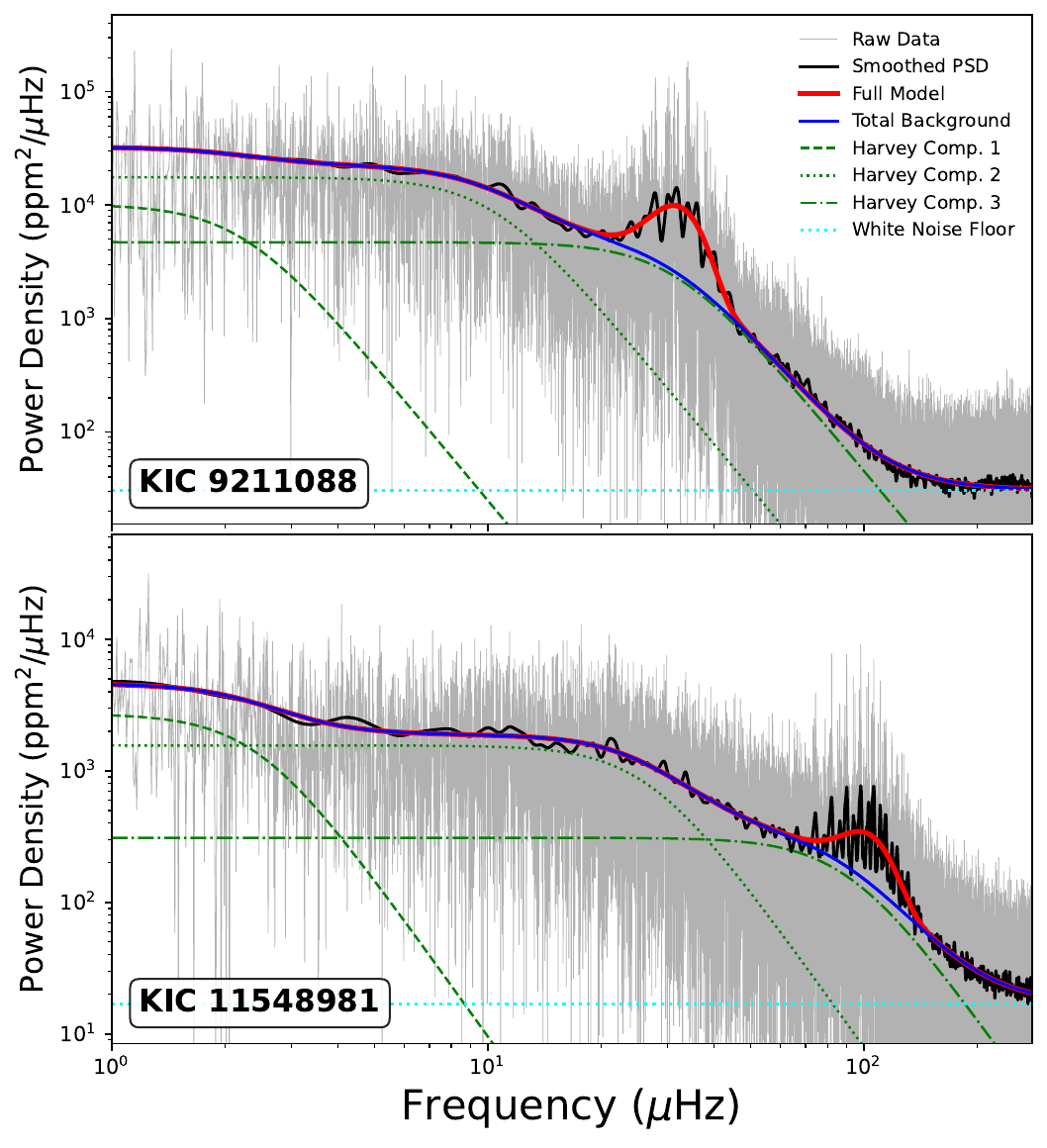}
	\end{center}
	\caption{Background fits for two representative stars. The gray line shows the observed power spectral density, while the black line represents the spectrum smoothed using a Gaussian filter. The solid red line shows the full model, including the background and the Gaussian oscillation envelope. The solid blue line represents the total background fit, the green lines show the three Harvey components, and the cyan dotted line indicates the white-noise level.}
	\label{fig:background_fit}
\end{figure}

The model parameters were determined sequentially. First,
the three Harvey-like components and the white-noise level were fitted
while excluding the frequency region containing the oscillation power
excess. The resulting background was then held fixed while the Gaussian
oscillation envelope was fitted to the smoothed PSD. The center of the
Gaussian was adopted as the frequency of maximum oscillation power,
$\nu_{\rm max}$. This procedure reduces the degeneracy between the
background components and the oscillation power excess while retaining
a complete model of the observed PSD.
The background model fitted to the power spectrum is shown in Figure~\ref{fig:background_fit}.

The stellar masses, radii, and surface gravities were subsequently
estimated from the standard asteroseismic scaling relations
\citep{kjeldsen1995},

\begin{align}
	\frac{R}{R_\odot} &=
	\left(\frac{\nu_{\rm max}}{\nu_{\rm max,\odot}}\right)
	\left(\frac{\Delta\nu}{\Delta\nu_\odot}\right)^{-2}
	\left(\frac{T_{\rm eff}}{T_{\rm eff,\odot}}\right)^{1/2},\\
	\frac{M}{M_\odot} &=
	\left(\frac{\nu_{\rm max}}{\nu_{\rm max,\odot}}\right)^3
	\left(\frac{\Delta\nu}{\Delta\nu_\odot}\right)^{-4}
	\left(\frac{T_{\rm eff}}{T_{\rm eff,\odot}}\right)^{3/2},\\
	\frac{g}{g_\odot} &=
	\left(\frac{\nu_{\rm max}}{\nu_{\rm max,\odot}}\right)
	\left(\frac{T_{\rm eff}}{T_{\rm eff,\odot}}\right)^{1/2},
\end{align}

\bigskip
\noindent where  
$\nu_{{\rm max}, \odot}
= 3090.0$ $\mu$Hz, $\Delta\nu_{\odot} = 135.1$ $\mu$Hz,
$T_{{\rm eff},\odot} = 5777.0$ K, $g_\odot = 4.438$, and the effective temperatures derived in Sect.~\ref{sec:calibrations}  were adopted in these relations.  

The derived global asteroseismic parameters for the red giant  stars are presented in Table~\ref{tab:seismic}.  
The quoted uncertainties in $\nu_{\rm max}$ and $\Delta\nu$ were estimated using 200 Monte Carlo realizations. In each realization, the PSD was resampled according to the statistical distribution of the power
spectrum and $\nu_{\rm max}$ and $\Delta\nu$ were re-determined using the same procedure as for the observed spectrum. Their uncertainties were taken as the standard deviations of the resulting distributions. Formal errors in mass, radius, and surface gravity were obtained by propagating those in $\nu_{\rm max}$, $\Delta\nu$, and $T_{\rm eff}$.
As can be noted, the errors in surface gravities are typically 0.004--0.005 dex. Such small uncertainties are expected from the strong dependence of surface gravity on $\nu_{\rm max}$, $g\propto\nu_{\rm max}T_{\rm eff}^{1/2}$, and are comparable to those reported for red giants in the SAGA catalog, where typical lower and upper uncertainties in $\log g$ are of the order of a few thousandths of a dex.

\begin{longtable}{rrcrcrrrrrrrcc}
	\caption{Global asteroseismic parameters for red giant stars.}\label{tab:seismic} \\
	\hline 
	KIC ID & $\nu_{\rm max}$ & $\sigma_{\nu_{\rm max}}¨$ & $\Delta 
	\nu$ & $\sigma_{\Delta \nu}$ & $M$ & $\sigma_{M}$ & $R$ & $\sigma_{R}$ & $log\,g$ & $\sigma_{log\,g}$ \\
	
	  &($\mu$Hz)&($\mu$Hz)&($\mu$Hz)&($\mu$Hz)& ($M_{\odot}$)& ($M_{\odot}$)& ($R_{\odot}$)& ($R_{\odot})$& (dex) &(dex) \\  
\hline
757137 & 29.9 & 0.2 & 3.45 & 0.01 & 2.19 & 0.06 & 14.98 & 0.14 & 2.428 & 0.004 \\
2161731 & 102.0 & 0.6 & 8.17 & 0.01 & 2.07 & 0.05 & 8.27 & 0.07 & 2.919 & 0.004 \\
2305930 & 26.9 & 0.3 & 3.98 & 0.02 & 0.64 & 0.03 & 9.02 & 0.15 & 2.332 & 0.005 \\
2309550 & 50.4 & 0.3 & 5.04 & 0.01 & 1.70 & 0.05 & 10.7 & 0.10 & 2.611 & 0.004 \\
2833697 & 31.5 & 0.3 & 4.16 & 0.01 & 1.15 & 0.04 & 10.65 & 0.14 & 2.443 & 0.006 \\
2969462 & 33.7 & 0.2 & 3.49 & 0.01 & 2.05 & 0.07 & 14.55 & 0.17 & 2.425 & 0.004 \\
3427850 & 27.2 & 0.2 & 3.49 & 0.01 & 1.11 & 0.03 & 11.86 & 0.12 & 2.336 & 0.004 \\
3443331 & 13.0 & 0.2 & 1.65 & 0.01 & 2.12 & 0.12 & 24.29 & 0.49 & 1.994 & 0.008 \\
3558705 & 48.3 & 0.3 & 4.66 & 0.01 & 2.01 & 0.07 & 11.90 & 0.13 & 2.589 & 0.005 \\
3730953 & 48.3 & 0.2 & 5.0 & 0.01 & 1.69 & 0.05 & 10.74 & 0.10 & 2.605 & 0.004 \\
3748585 & 46.9 & 0.3 & 4.65 & 0.01 & 1.72 & 0.05 & 11.31 & 0.12 & 2.565 & 0.004 \\
3833399 & 38.6 & 0.2 & 4.21 & 0.01 & 1.50 & 0.04 & 11.56 & 0.10 & 2.489 & 0.003 \\
3854605 & 39.2 & 0.3 & 4.21 & 0.01 & 1.51 & 0.05 & 11.59 & 0.12 & 2.489 & 0.004 \\
3936801 & 223.1 & 0.4 & 15.17 & 0.01 & 1.77 & 0.06 & 5.19 & 0.06 & 3.254 & 0.005 \\
4038624 & 58.8 & 0.3 & 5.52 & 0.01 & 1.94 & 0.07 & 10.52 & 0.12 & 2.682 & 0.005 \\
4135680 & 8.0 & 0.1 & 1.36 & 0.01 & 1.06 & 0.05 & 21.84 & 0.42 & 1.784 & 0.006 \\
4283484 & 30.1 & 0.2 & 4.35 & 0.02 & 0.66 & 0.03 & 8.60 & 0.14 & 2.389 & 0.006 \\
4348593 & 67.3 & 0.7 & 5.63 & 0.01 & 2.77 & 0.10 & 11.68 & 0.14 & 2.745 & 0.005 \\
4457200 & 44.7 & 0.2 & 4.49 & 0.01 & 1.76 & 0.05 & 11.68 & 0.12 & 2.549 & 0.004 \\
4479314 & 19.8 & 0.2 & 2.24 & 0.01 & 2.26 & 0.10 & 20.17 & 0.31 & 2.183 & 0.006 \\
4480583 & 42.6 & 0.6 & 4.5 & 0.06 & 1.53 & 0.11 & 11.13 & 0.33 & 2.529 & 0.007 \\
4571402 & 39.0 & 0.2 & 4.43 & 0.01 & 1.34 & 0.04 & 10.77 & 0.12 & 2.502 & 0.004 \\
4643888 & 51.0 & 0.2 & 5.01 & 0.01 & 1.78 & 0.04 & 10.88 & 0.08 & 2.614 & 0.003 \\
4667911 & 37.1 & 0.3 & 4.36 & 0.01 & 1.14 & 0.03 & 10.31 & 0.11 & 2.469 & 0.004 \\
4833670 & 111.5 & 0.6 & 8.65 & 0.01 & 2.18 & 0.06 & 8.10 & 0.08 & 2.959 & 0.004 \\
4845272 & 31.0 & 0.2 & 3.61 & 0.01 & 1.24 & 0.03 & 12.02 & 0.11 & 2.372 & 0.004 \\
4934550 & 182.6 & 0.1 & 14.8 & 0.01 & 1.16 & 0.02 & 4.58 & 0.02 & 3.178 & 0.002 \\
4936438 & 116.7 & 0.6 & 9.68 & 0.01 & 1.57 & 0.04 & 6.74 & 0.05 & 2.977 & 0.003 \\
4989752 & 42.6 & 0.3 & 4.53 & 0.01 & 1.41 & 0.04 & 10.79 & 0.12 & 2.521 & 0.004 \\
5026399 & 112.1 & 0.8 & 8.65 & 0.01 & 2.33 & 0.08 & 8.28 & 0.09 & 2.969 & 0.005 \\
5121116 & 23.3 & 0.2 & 2.68 & 0.01 & 1.94 & 0.07 & 16.99 & 0.22 & 2.265 & 0.005 \\
5307747 & 83.0 & 0.4 & 6.93 & 0.01 & 2.18 & 0.07 & 9.39 & 0.11 & 2.831 & 0.005 \\
5631061 & 22.2 & 0.3 & 2.34 & 0.01 & 2.91 & 0.13 & 21.33 & 0.37 & 2.243 & 0.006 \\
5700368 & 36.4 & 0.2 & 4.16 & 0.01 & 1.32 & 0.04 & 11.17 & 0.13 & 2.464 & 0.005 \\
5701829 & 141.3 & 0.3 & 12.15 & 0.01 & 1.10 & 0.02 & 5.14 & 0.03 & 3.057 & 0.003 \\
5795626 & 38.1 & 0.2 & 4.58 & 0.01 & 1.10 & 0.03 & 9.85 & 0.09 & 2.492 & 0.003 \\
6127825 & 108.8 & 0.7 & 8.48 & 0.01 & 2.22 & 0.05 & 8.26 & 0.07 & 2.951 & 0.003 \\
6181933 & 67.5 & 0.3 & 6.21 & 0.01 & 1.82 & 0.06 & 9.52 & 0.10& 2.741 & 0.004 \\
6231193 & 31.7 & 0.2 & 3.91 & 0.01 & 1.05 & 0.03 & 10.79 & 0.11 & 2.394 & 0.004 \\
6283995 & 198.0 & 0.4 & 14.78 & 0.01 & 1.37 & 0.04 & 4.86 & 0.05 & 3.202 & 0.004 \\
7175316 & 45.0 & 0.6 & 4.02 & 0.02 & 3.01 & 0.14 & 15.02 & 0.26 & 2.562 & 0.006 \\
7187063 & 29.7 & 0.2 & 3.98 & 0.01 & 0.93 & 0.05 & 10.25 & 0.18 & 2.388 & 0.007 \\
7191496 & 16.1 & 0.1 & 2.5 & 0.01 & 0.96 & 0.03 & 14.07 & 0.17 & 2.122 & 0.004 \\
7286856 & 201.1 & 0.6 & 14.7 & 0.01 & 1.46 & 0.06 & 4.97 & 0.07 & 3.208 & 0.006 \\
7340724 & 131.1 & 0.6 & 10.59 & 0.01 & 1.51 & 0.04 & 6.26 & 0.05 & 3.023 & 0.003 \\
7514596 & 31.8 & 0.3 & 4.03 & 0.01 & 0.96 & 0.04 & 10.28 & 0.13 & 2.398 & 0.005 \\
7668613 & 38.1 & 0.2 & 4.31 & 0.01 & 1.34 & 0.04 & 10.96 & 0.12 & 2.484 & 0.005 \\
7730752 & 61.7 & 0.3 & 5.81 & 0.01 & 1.81 & 0.10 & 9.93 & 0.19 & 2.702 & 0.008 \\
8210100 & 42.7 & 0.2 & 4.43 & 0.01 & 1.61 & 0.05 & 11.45 & 0.13 & 2.528 & 0.005 \\
8280417 & 35.4 & 0.2 & 4.2 & 0.01 & 1.17 & 0.04 & 10.66 & 0.13 & 2.450 & 0.005 \\
8312388 & 44.6 & 0.3 & 4.45 & 0.01 & 1.91 & 0.07 & 12.08 & 0.14 & 2.555 & 0.005 \\
8390838 & 52.1 & 0.3 & 5.09 & 0.01 & 1.81 & 0.06 & 10.84 & 0.12 & 2.626 & 0.005 \\
8539201 & 33.6 & 0.2 & 3.83 & 0.01 & 1.57 & 0.04 & 12.51 & 0.12 & 2.439 & 0.004 \\
8823410 & 54.4 & 0.3 & 5.35 & 0.01 & 1.62 & 0.09 & 10.11 & 0.20 & 2.639 & 0.008 \\
9003096 & 69.2 & 0.3 & 6.31 & 0.01 & 1.79 & 0.06 & 9.36 & 0.10 & 2.749 & 0.005 \\
9071940 & 38.2 & 0.2 & 4.33 & 0.01 & 1.13 & 0.09 & 10.33 & 0.28 & 2.464 & 0.012 \\
9211088 & 31.7 & 0.2 & 3.94 & 0.01 & 1.16 & 0.04 & 11.11 & 0.14 & 2.413 & 0.005 \\
9813798 & 47.0 & 0.3 & 4.58 & 0.01 & 2.01 & 0.05 & 12.04 & 0.11 & 2.579 & 0.003 \\
9897634 & 68.8 & 0.3 & 6.02 & 0.01 & 2.11 & 0.04 & 10.20 & 0.07 & 2.745 & 0.003 \\
10007364 & 35.3 & 0.3 & 4.05 & 0.01 & 1.28 & 0.04 & 11.25 & 0.12 & 2.442 & 0.004 \\
10198347 & 41.5 & 0.6 & 4.35 & 0.49 & 1.73 & 0.79 & 11.86 & 2.70 & 2.528 & 0.007 \\
10266802 & 75.9 & 0.4 & 6.41 & 0.01 & 2.39 & 0.07 & 10.19 & 0.11 & 2.799 & 0.004 \\
10333063 & 36.5 & 0.2 & 4.05 & 0.01 & 1.65 & 0.05 & 12.25 & 0.14 & 2.479 & 0.005 \\
10513837 & 195.8 & 0.4 & 14.97 & 0.01 & 1.32 & 0.03 & 4.76 & 0.04 & 3.204 & 0.004 \\
10790415 & 37.8 & 0.3 & 4.24 & 0.01 & 1.36 & 0.04 & 11.13 & 0.12 & 2.478 & 0.004 \\
10918449 & 41.8 & 0.3 & 3.8 & 0.01 & 3.07 & 0.09 & 15.71 & 0.18 & 2.533 & 0.004 \\
10936814 & 37.5 & 0.2 & 4.37 & 0.01 & 1.17 & 0.04 & 10.39 & 0.11 & 2.474 & 0.004 \\
11234897 & 60.5 & 0.5 & 5.23 & 0.01 & 2.73 & 0.08 & 12.20 & 0.12 & 2.701 & 0.004 \\
11235672 & 73.9 & 0.4 & 5.93 & 0.01 & 2.97 & 0.06 & 11.54 & 0.08 & 2.786 & 0.003 \\
11251115 & 55.3 & 0.4 & 5.12 & 0.01 & 2.1 & 0.06 & 11.34 & 0.11 & 2.651 & 0.004 \\
11338492 & 23.8 & 0.2 & 2.79 & 0.01 & 1.72 & 0.05 & 15.9 & 0.18 & 2.271 & 0.004 \\
11352446 & 91.3 & 0.5 & 7.64 & 0.01 & 1.88 & 0.05 & 8.38 & 0.08 & 2.866 & 0.004 \\
11547584 & 48.3 & 0.2 & 4.81 & 0.01 & 1.8 & 0.05 & 11.24 & 0.10 & 2.592 & 0.004 \\
11548981 & 101.1 & 0.5 & 8.12 & 0.01 & 2.16 & 0.07 & 8.42 & 0.10 & 2.921 & 0.005 \\
11600024 & 36.3 & 0.3 & 4.31 & 0.01 & 1.15 & 0.03 & 10.41 & 0.10 & 2.462 & 0.004 \\
11600442 & 69.5 & 0.3 & 6.19 & 0.01 & 2.00 & 0.06 & 9.84 & 0.10 & 2.754 & 0.004 \\
11603064 & 16.3 & 0.2 & 1.94 & 0.01 & 2.29 & 0.11 & 22.28 & 0.38 & 2.102 & 0.006 \\
11618103 & 106.9 & 0.3 & 9.26 & 0.01 & 1.41 & 0.03 & 6.69 & 0.05 & 2.936 & 0.003 \\
11651229 & 45.1 & 0.3 & 4.37 & 0.01 & 2.14 & 0.05 & 12.70 & 0.11 & 2.561 & 0.003 \\
\hline
\end{longtable}

\begin{figure}[!hb]
	\begin{center}
		\includegraphics[scale=0.35]
		{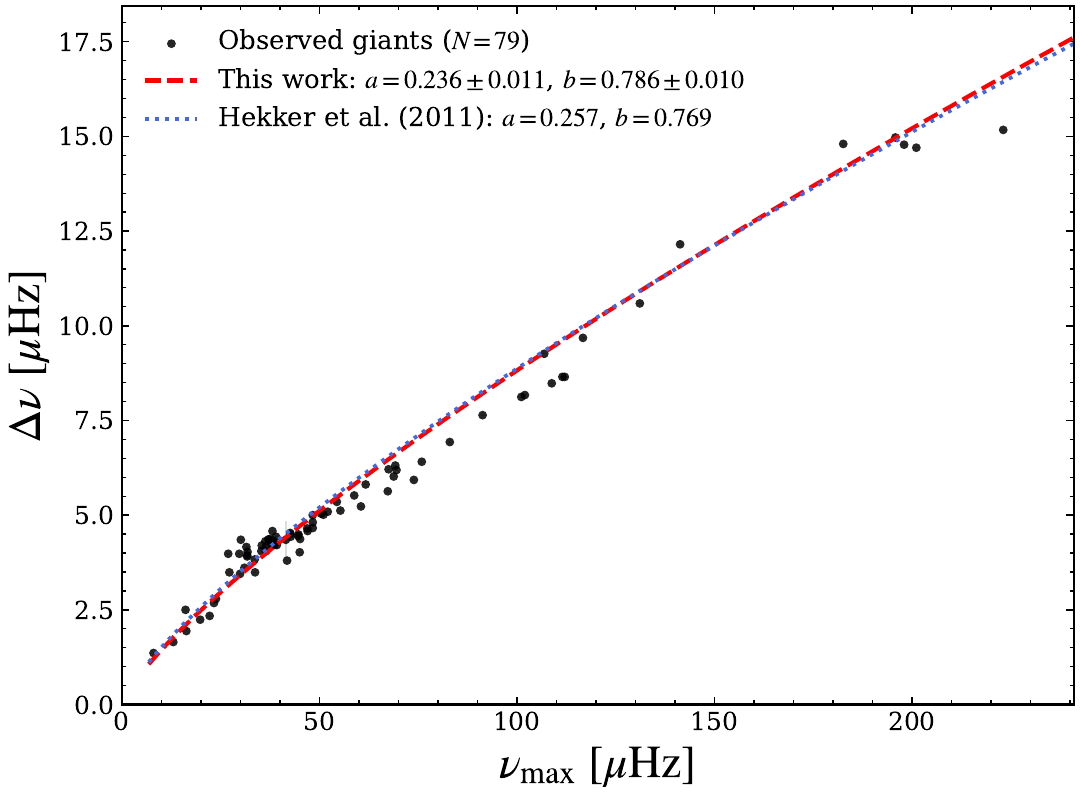}
	\end{center}
	\caption{$\Delta\nu$ versus $\nu_{\rm max}$ for the 79 red giants with global asteroseismic measurements. Black points represent the observed values, while the red dashed line shows the power-law fit to our sample, $\Delta\nu=a\nu_{\rm max}^{b}$, with $a=0.236\pm0.011$ and $b=0.786\pm0.010$. The blue dotted line represents the relation obtained by \citet{hekker2011}.}
	\label{fig:numax_dnu}
\end{figure}

The global asteroseismic parameters derived for the red giants (Table~\ref{tab:seismic}) follow the well-established correlation between the frequency of maximum oscillation power, $\nu_{\rm max}$, and the large frequency separation, $\Delta\nu$ \citep{mosser2010,hekker2011}. As shown in Fig.~\ref{fig:numax_dnu}, we fitted the observed relation with a power law of the form $\Delta\nu=a\nu_{\rm max}^{b}$, obtaining $a=0.236\pm0.011$ and $b=0.786\pm0.010$. These coefficients are in reasonable agreement with the values reported by \citet{hekker2011}, supporting the consistency of our asteroseismic measurements with the expected behavior of red giant oscillations.

\begin{figure*}
	\begin{center}
		\includegraphics[scale=0.35]
		{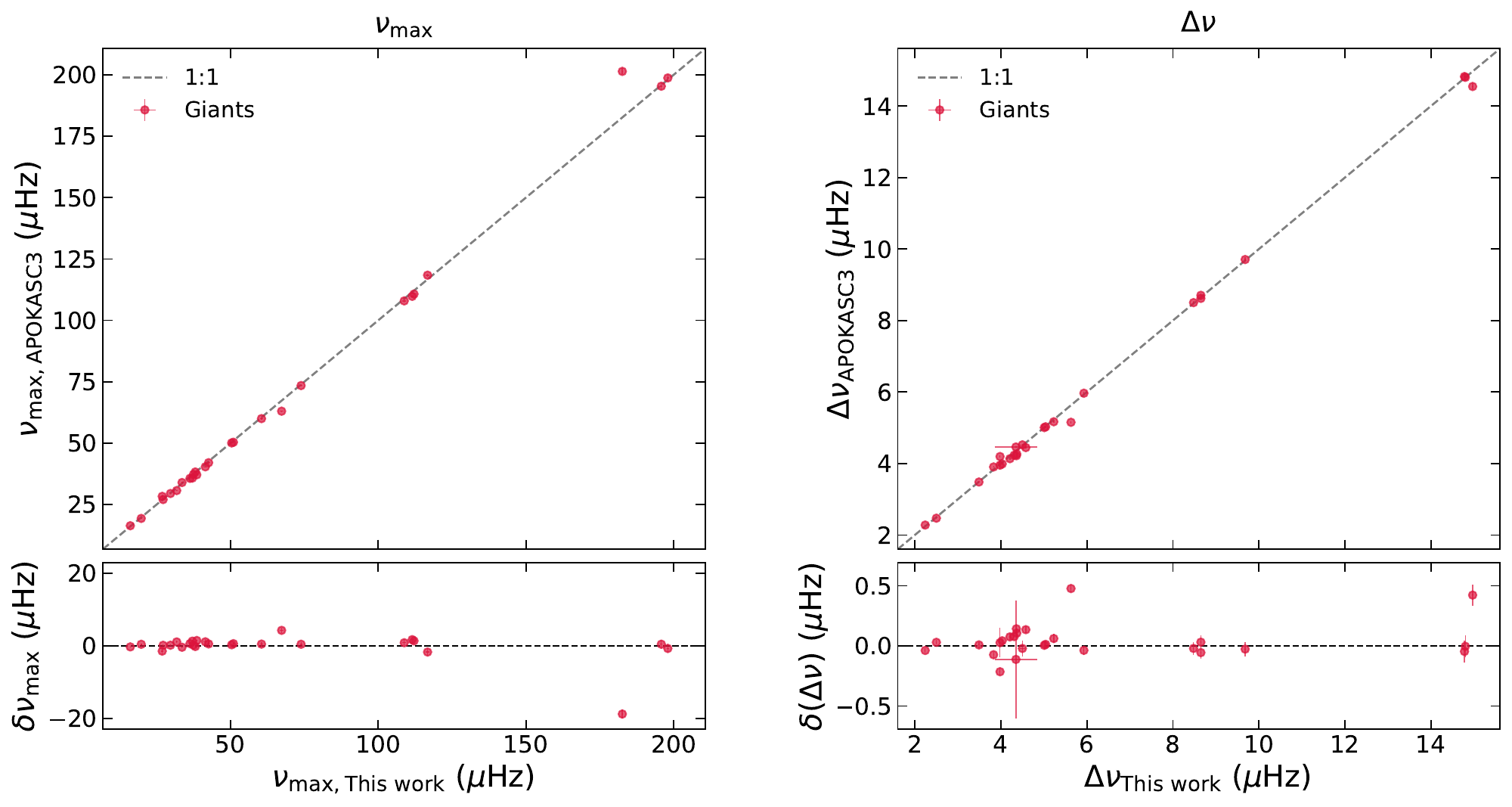}
	\end{center}
	\caption{Comparison of the global asteroseismic parameters $\nu_{\rm max}$ (left) and $\Delta\nu$ (right) 
		derived in this work with those reported in APOKASC-3 for stars with valid measurements in both samples. 
		The upper panels show the direct comparisons, with dashed lines indicating the one-to-one relations. 
		The lower panels show the differences (this work minus APOKASC-3), with horizontal dashed lines indicating zero difference. 
		Error bars represent the uncertainties in both measurements, propagated in quadrature for the differences. 
		In most cases, the error bars are smaller than or comparable to the symbol sizes and are therefore difficult to distinguish.}
	\label{fig:comp_numax_dnu}
\end{figure*}

{ In addition, Fig.~\ref{fig:comp_numax_dnu} presents a direct comparison of our independently derived $\nu_{\rm max}$ and $\Delta\nu$ 
	measurements with those reported in APOKASC-3. The overall agreement is good, with mean differences (this work minus APOKASC-3) of $-0.210\pm3.945$ $\mu$Hz for 
	$\nu_{\rm max}$ and $+0.039\pm0.143$ $\mu$Hz for $\Delta\nu$, where the quoted dispersions represent the standard deviations of the differences. 
	The corresponding median differences are $+0.464$ and $+0.011$ $\mu$Hz, respectively. 
	These results provide additional support for the reliability of our independently derived global asteroseismic parameters.}

\section{Bayesian inference of stellar parameters from isochrones}
\label{sec:parsec}

A fundamental difficulty in deriving stellar parameters from  isochrones  is that models with different mass, age and evolutionary stages may reproduce similar observed stellar properties. Bayesian inference provides a statistical framework for addressing this degeneracy by combining the likelihood of the observations with prior information on stellar populations \citep[e.g.,][]{pont2004,takeda2007,david2015}. 
Rather than relying on a single best-fitting model, the method assigns probabilities to the evolutionary solutions compatible with the observations, allowing the stellar parameters and their uncertainties to be estimated from the resulting posterior distributions.

We applied a Bayesian approach to determine the surface gravities ($\log g$), which were not obtained directly from our Strömgren–Crawford photometric analysis, and to further constraint other fundamental stellar parameters. The analysis combines photometric effective temperatures and metallicities derived in Sect.~\ref{sec:calibrations} with absolute visual magnitudes calculated using \textit{Gaia} DR3 distances or, when available, global asteroseismic parameters. 

The reference grid was constructed from
PARSEC stellar isochrones  
\citep{bressan2012}\footnote{\url{https://stev.oapd.inaf.it/cgi-bin/cmd}}, covering initial masses from $0.09$ to $2.40\,M_\odot$, ages from 1 to 10 Gyr,  global metallicities from $[\mathrm{M/H}]=-2.19$ to $+0.40$ dex. The isochrones were sampled at intervals of 0.10 in $\log_{10}(\mathrm{Age/yr})$ and approximately 0.10 dex in [M/H]. To improve the numerical resolution,  we interpolated in initial mass along each individual isochrone, subdividing the intervals between consecutive stellar models into ten equal segments. Interpolation between isochrones of different ages or metallicities was not performed, following the methodological considerations of \citet{david2015} regarding potential interpolation errors arising from the nonlinear evolution of stellar properties with age.

The PARSEC models adopt a scaled-solar heavy-element composition, with a reference solar metallicity of $Z_\odot=0.0152$. The initial helium abundance follows the enrichment relation $Y=0.2485+1.78Z$, while convective mixing and overshooting are treated according to the prescriptions described by \citet{bressan2012}. Mass loss along the red giant branch is modeled using the Reimers prescription with $\eta_{\rm Reimers}=0.2$. Further details of the input physics and evolutionary calculations are given by \citet{bressan2012}.

To ensure consistency between the observational metallicities and those adopted in the PARSEC isochrones, we converted the photometric iron abundances, $[\mathrm{Fe/H}]$, derived in Sect.~\ref{sec:calibrations} into global metallicities, $[\mathrm{M/H}]$, following the $\alpha$-enhancement prescription of \citet{pagel1995}. The converted values and their propagated uncertainties were then used as metallicity constraints in the Bayesian analysis.

For each model, the posterior probability was evaluated according to Bayes' theorem,

\begin{equation}
	p(\mathbf{x}|\mathbf{y}) \propto
	p(\mathbf{y}|\mathbf{x})\,p(\mathbf{x}),
\end{equation}

\noindent where $\mathbf{x}$ represents the model parameters and $\mathbf{y}$ the observed quantities. Assuming independent
Gaussian observational errors, the likelihood was expressed as

\begin{equation}
	\mathcal{L} \propto \exp\left(-\frac{\chi^2}{2}\right).
\end{equation}

The prior weights incorporated the Kroupa initial mass function (IMF; \citealt{kroupa2001}) and the 
volume represented by each model in initial-mass, age, and metallicity space:

\begin{equation}
	w_{\rm prior} \propto
	\xi(M_{\rm ini})\,
	\Delta M_{\rm ini}\,
	\Delta t\,
	\Delta [{\rm M/H}],
\end{equation}

\noindent where $\xi(M_{\rm ini})$ is the IMF. We assumed a uniform star-formation rate in linear age and a flat metallicity prior. 
The grid-volume factors account for the nonuniform sampling of parameter space, preventing densely sampled regions from receiving artificially enhanced posterior weights.

For dwarfs and giants without asteroseismic measurements, the likelihood was constrained by $T_{\rm eff}$, [M/H], and the absolute visual magnitude $M_V$, calculated from the dereddened apparent magnitude $V_0$ and the adopted \textit{Gaia} DR3 distance. The $\chi^2$ was then evaluated as

\begin{equation}
	\begin{split}
	    \chi^2 = {}& 
	    \left(
       	\frac{T_{\rm eff}^{\rm mod}-T_{\rm eff}^{\rm obs}}
     	{\sigma_{T_{\rm eff}}}
    	\right)^2
     	+
     	\left(
    	\frac{[{\rm M/H}]^{\rm mod}-[{\rm M/H}]^{\rm obs}}
    	{\sigma_{[{\rm M/H}]}}
	    \right)^2\\
     	&  \hspace{2.0cm} +
   	\left(
	    \frac{M_V^{\rm mod}-M_V^{\rm obs}}
	    {\sigma_{M_V}}
     	\right)^2 .
    \end{split}
\end{equation}

We note that the Strömgren indices were not included as separate likelihood terms, since they had already been used to derive the photometric effective temperatures and metallicities.

For giants with reliable global asteroseismic measurements, $M_V$ was replaced by the frequency of maximum oscillation power, $\nu_{\rm max}$, and the large frequency separation, $\Delta\nu$. The model predictions were calculated from the PARSEC masses, radii, effective temperatures, and surface gravities using the standard asteroseismic scaling relations. The corresponding $\chi^2$ was expressed as

\begin{equation}
	\begin{split}
		\chi^2 ={}&
		\left(
		\frac{T_{\rm eff}^{\rm mod}-T_{\rm eff}^{\rm obs}}
		{\sigma_{T_{\rm eff}}}
		\right)^2
		+
		\left(
		\frac{[{\rm M/H}]^{\rm mod}-[{\rm M/H}]^{\rm obs}}
		{\sigma_{[{\rm M/H}]}}
		\right)^2 \\
		&+
		\left(
		\frac{\nu_{\rm max}^{\rm mod}-\nu_{\rm max}^{\rm obs}}
		{\sigma_{\nu_{\rm max}}}
		\right)^2
		+
		\left(
		\frac{\Delta\nu^{\rm mod}-\Delta\nu^{\rm obs}}
		{\sigma_{\Delta\nu}}
		\right)^2 .
	\end{split}
\end{equation}

To avoid double counting the asteroseismic information, masses, radii, and surface gravities independently derived from $\nu_{\rm max}$ and $\Delta\nu$ were not included as additional likelihood constraints.

The posterior probability distributions were marginalized over the grid to obtain the parameters of interest. We adopted the posterior medians as the central estimates and the 16th and 84th percentiles to define asymmetric credible intervals. Marginal distributions were calculated for $T_{\rm eff}$, [M/H], current stellar mass, initial mass, radius, $\log g$, age, and luminosity.

To assess the sensitivity of the inferred stellar parameters to the numerical resolution and interpolation of the model grid, we repeated the Bayesian analysis using an alternative three-dimensional grid constructed by linear interpolation in initial mass, age, and metallicity. The interpolated grid adopted steps of 0.05 in $\log_{10}(\mathrm{Age/yr})$ and 0.05 dex in [M/H]. Both grids yielded generally consistent fundamental parameters.
Table~\ref{tab:stellar_params} presents the stellar parameters derived from the Bayesian analysis. The complete results are provided in machine-readable form.

\begin{figure*}
	\begin{center}
		\includegraphics[scale=0.35]
		{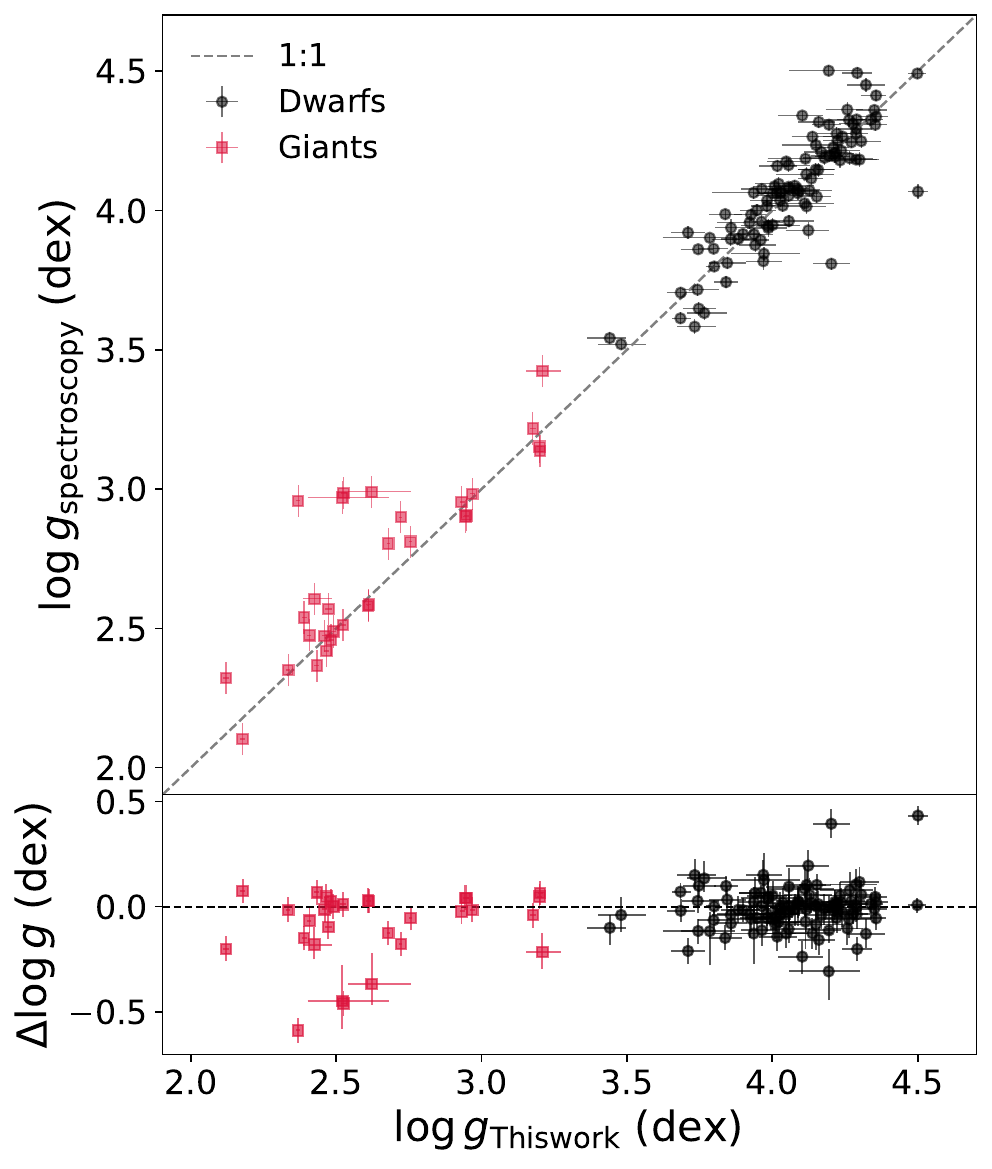}
	\end{center}
	\caption{Comparison of the surface gravities derived from the Bayesian analysis in this work with those reported by APOGEE-2 for 104 dwarfs (black circles) and APOKASC-3 for 30 giants (red squares). The upper panel shows the comparison with the one-to-one relation (dashed line), while the lower panel shows the differences $\Delta\log g=\log g_{\rm Bayes}-\log g_{\rm spectroscopy}$. Error bars represent the asymmetric uncertainties from the Bayesian analysis and the reported spectroscopic uncertainties, propagated in quadrature for the residuals. The horizontal dashed line in the lower panel indicates zero difference.{The surface gravities derived in our work for dwarf stars were obtained through Bayesian isochrone 
			fitting using only classical observational constraints, namely $T_{\rm eff}$, [Fe/H], and $M_V$. No asteroseismic parameters were determined or 
			used for dwarf stars. In contrast, for red giants with detected solar-like oscillations, 
			the additional constraints $\nu_{\max}$ and $\Delta\nu$ were incorporated into the Bayesian analysis. 
	 } }
	\label{fig:comp_logg}
\end{figure*}

The surface gravities derived from the Bayesian analysis were compared with those reported by APOGEE-2 and APOKASC-3 for the same  dwarfs and  giants considered in the previous comparisons (Fig.~\ref{fig:comp_logg}). {The surface gravities adopted from APOGEE-2 DR17 for dwarf stars correspond to the calibrated ASPCAP values (LOGG), rather than individual asteroseismic determinations. In contrast, the surface gravities derived in this work for dwarf stars were obtained through Bayesian isochrone fitting using only classical observational constraints, namely $T_{\rm eff}$, [Fe/H], and $M_V$, without incorporating asteroseismic information.}  Overall, the agreement is good, with a mean difference (this work minus spectroscopy) of $\Delta\log g=-0.027\pm0.127$ dex and a median of $-0.013$ dex for the 134 stars. The two sets of gravities are strongly correlated ($r=0.980$; upper panel of Fig.~\ref{fig:comp_logg}), although this correlation is partly driven by the wide range of surface gravities covered by the combined dwarf and giant samples.

For the dwarfs in common with APOGEE-2, the mean difference is $\Delta\log g=-0.008\pm0.104$ dex (median $-0.010$ dex), indicating excellent agreement between the Bayesian and spectroscopic gravities, with no appreciable systematic offset. This result is noteworthy given the systematic difference of approximately $+230$ K found between our photometric effective temperatures and those of APOGEE-2.

For the giants in common with APOKASC-3, the mean difference is $\Delta\log g=-0.092\pm0.174$ dex, whereas the median is only $-0.019$ dex. The larger dispersion and the difference between the mean and median are associated with a small number of stars showing substantially lower Bayesian gravities, as illustrated in the residuals (lower panel of Fig.~\ref{fig:comp_logg}). Most giants, however, lie relatively close to the one-to-one relation. 

{ We also compared the stellar masses, radii, and ages derived from our Bayesian analysis
	with those reported in APOKASC-3 for red giants with valid parameters in both samples.
	The mean differences (this work minus APOKASC-3) are $-0.174\pm0.259\,M_\odot$, $-0.388\pm0.586\,R_\odot$, and $+0.272\pm1.489$ Gyr
	for mass, radius, and age, respectively, where the quoted dispersions represent the standard deviations of the differences.
	The corresponding median differences are $-0.097\,M_\odot$, $-0.303\,R_\odot$, and $+0.254$ Gyr. The differences may reflect the distinct evolutionary models,
	adopted observational constraints, and inference procedures employed in the two studies.  It is worth noting that \citet{pinso2025} employed asteroseismic methods, including techniques based on dipole mixed modes, to distinguish between RGB and RC stars, providing additional information 
	on their evolutionary states that was not included in our analysis.}

\begin{table*}[!t]
	\caption{Derived fundamental parameters for the stellar sample.}
	\label{tab:stellar_params}
	\centering
	\setlength{\tabcolsep}{3pt}
	\small
	\begin{tabular}{rlrccrc}
		\hline
		\hline
		KIC ID & $T_{\mathrm{eff}}$ & $[\mathrm{M/H}]$ &
		$\log g$ & Mass & Radius & Lum. Class \\
		& (K) & (dex) & (dex) &
		($M_{\odot}$) & ($R_{\odot}$) & \\
		\hline
		1430163 & $6735_{-147}^{+134}$ & $-0.10_{-0.20}^{+0.10}$ & $4.216_{-0.068}^{+0.050}$ & $1.30_{-0.07}^{+0.07}$ & $1.47_{-0.10}^{+0.12}$ & Dwarf \\
		1435467 & $6252_{-358}^{+294}$ & $0.20_{-0.10}^{+0.10}$ & $4.111_{-0.131}^{+0.102}$ & $1.34_{-0.10}^{+0.05}$ & $1.68_{-0.17}^{+0.24}$ & Dwarf \\
		2161731 & $4928_{-10}^{+61}$ & $0.10_{-0.10}^{+0.00}$ & $2.912_{-0.006}^{+0.007}$ & $1.99_{-0.03}^{+0.01}$ & $8.15_{-0.12}^{+0.08}$ & Giant \\
		2305930 & $4751_{-42}^{+10}$ & $-0.60_{-0.00}^{+0.10}$ & $2.369_{-0.006}^{+0.007}$ & $0.91_{-0.00}^{+0.02}$ & $10.35_{-0.06}^{+0.11}$ & Giant \\
		2309550 & $4874_{-50}^{+27}$ & $-0.10_{-0.00}^{+0.10}$ & $2.610_{-0.005}^{+0.004}$ & $1.66_{-0.01}^{+0.05}$ & $10.57_{-0.04}^{+0.12}$ & Giant \\
		2833697 & $5365_{-49}^{+50}$ & $-1.10_{-0.10}^{+0.10}$ & $2.434_{-0.012}^{+0.011}$ & $1.10_{-0.09}^{+0.10}$ & $10.46_{-0.41}^{+0.37}$ & Giant \\
		2837475 & $6716_{-135}^{+145}$ & $0.0_{-0.10}^{+0.20}$ & $4.153_{-0.060}^{+0.048}$ & $1.39_{-0.06}^{+0.05}$ & $1.63_{-0.09}^{+0.11}$ & Dwarf \\
		2969462 & $4662_{-47}^{+45}$ & $0.0_{-0.10}^{+0.10}$ & $2.416_{-0.009}^{+0.009}$ & $1.91_{-0.15}^{+0.12}$ & $14.11_{-0.39}^{+0.36}$ & Giant \\
		3123191 & $6342_{-106}^{+164}$ & $-0.1_{-0.10}^{+0.20}$ & $4.149_{-0.062}^{+0.065}$ & $1.21_{-0.08}^{+0.07}$ & $1.53_{-0.10}^{+0.09}$ & Dwarf \\
		3221671 & $6605_{-195}^{+138}$ & $0.0_{-0.10}^{+0.10}$ & $4.196_{-0.064}^{+0.046}$ & $1.33_{-0.06}^{+0.05}$ & $1.52_{-0.08}^{+0.10}$ & Dwarf \\
		3223000 & $6332_{-134}^{+116}$ & $-0.20_{-0.10}^{+0.10}$ & $4.289_{-0.063}^{+0.054}$ & $1.09_{-0.06}^{+0.06}$ & $1.24_{-0.05}^{+0.07}$ & Dwarf \\
		3236382 & $6870_{-127}^{+82}$ & $-0.10_{-0.20}^{+0.20}$ & $4.134_{-0.045}^{+0.041}$ & $1.38_{-0.07}^{+0.06}$ & $1.66_{-0.06}^{+0.07}$ & Dwarf \\
		3241299 & $6482_{-105}^{+138}$ & $0.0_{-0.10}^{+0.20}$ & $3.776_{-0.034}^{+0.059}$ & $1.60_{-0.03}^{+0.11}$ & $2.70_{-0.11}^{+0.11}$ & Dwarf \\
		3424541 & $6283_{-134}^{+132}$ & $0.20_{-0.10}^{+0.10}$ & $3.798_{-0.053}^{+0.051}$ & $1.62_{-0.07}^{+0.04}$ & $2.65_{-0.16}^{+0.16}$ & Dwarf \\
		3427720 & $6109_{-114}^{+98}$ & $0.0_{-0.10}^{+0.10}$ & $4.354_{-0.061}^{+0.042}$ & $1.09_{-0.06}^{+0.05}$ & $1.15_{-0.05}^{+0.06}$ & Dwarf \\
		\hline
	\end{tabular}
	\tablecomments{The full version of this table is published
		electronically in machine-readable format. A subset is
		displayed here for illustration purposes.}
\end{table*}

\begin{figure}
	\begin{center}
		\includegraphics[scale=0.35]
		{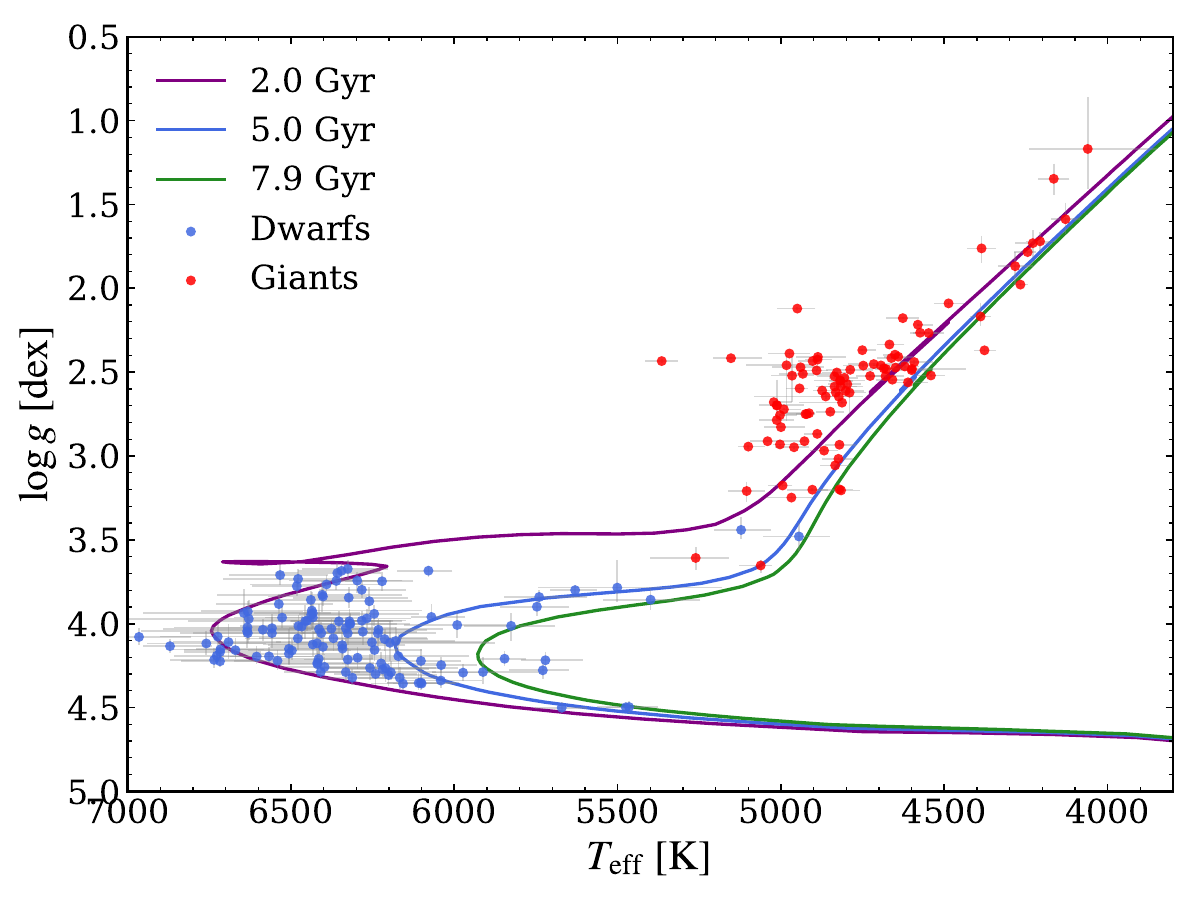}
	\end{center}
	\caption{Kiel diagram showing the effective temperatures and surface gravities derived from the Bayesian analysis for dwarfs and subgiants (blue circles) and  giants (red circles). The solid lines represent PARSEC isochrones at ages of 2.0, 5.0, and 7.9 Gyr, for approximately solar metallicity. Error bars indicate the asymmetric 16th–84th percentile credible intervals.  }
	\label{fig:hr}
\end{figure}

Figure~\ref{fig:hr} presents the Kiel diagram ($\log g$  versus $T_{\mathrm{eff}}$) for all sample. Three representative PARSEC isochrones with solar metallicity ([M/H] = 0.0 dex) and ages of 2.0, 5.0, and 7.9 Gyr are overplotted to illustrate the main evolutionary sequences.

Most of the dwarf population is concentrated at effective temperatures between approximately 6100 and 6700 K and surface gravities between 3.7 and 4.4 dex, consistent with the main-sequence region. However, several stars extend toward lower effective temperatures and surface gravities, occupying regions associated with the main-sequence turnoff and subgiant branch. In particular, objects with temperatures of approximately 5500–6100 K and log g values between 3.5 and 4.0 dex may represent evolved stars transitioning from the main sequence toward the red giant branch. This suggests that the dwarf sample may contain a fraction of subgiant stars.

The giant population is primarily distributed between 4500 and 5100 K, with surface gravities ranging from approximately 2.3 to 3.2 dex. A prominent concentration is observed around 4700–5000 K and log g = 2.4–2.8 dex. The distribution of these stars is consistent with an evolved population that may include both red giant branch and core-helium-burning stars.


\section{Summary and conclusions}
\label{sec:conclusions}

We obtained Str\"omgren--Crawford $uvby$--H$\beta$ photoelectric photometry for a sample of relatively bright ($Kp < 12$ mag) FGK-type stars in the \textit{Kepler} 
field. For most of these stars, Str\"omgren photometry is presented for the first time. Interstellar reddening was estimated using the Moon calibrations and 
the Bayestar19 three-dimensional dust maps. We distinguished between dwarf and giant stars and determined their metallicities, [Fe/H], and 
effective temperatures, $T_{\rm eff}$, using empirical photometric calibrations.

The derived atmospheric parameters were compared with those reported by SAGA, APOGEE-2, and APOKASC-3. 
For the 22 stars in common with SAGA, our photometric metallicities and effective temperatures show reasonable overall agreement with the SAGA determinations, 
although the temperature differences exhibit distinct behavior for dwarfs and giants. For the 104 dwarfs in common with APOGEE-2, 
our metallicities agree well with the spectroscopic values, whereas our effective temperatures are systematically higher by approximately 230 K. 
For the 30 giants in common with APOKASC-3, our metallicities are systematically lower by approximately 0.24 dex, 
while the effective temperatures show relatively good agreement, with our values being lower by approximately 70 K. 
The metallicity offset found for these giants may be related to the metallicity-dependent differences 
reported by \citet{casagrande2014} in their comparison with APOGEE. However, the limited size of our 
common sample and differences in the spectroscopic data releases prevent a direct assessment of this possibility.

We also analyzed the \textit{Kepler} light curves of the red giants exhibiting solar-like oscillations 
to determine the global asteroseismic parameters, including the frequency of maximum oscillation power, $\nu_{\max}$, 
and the large frequency separation, $\Delta\nu$.

A Bayesian isochrone-fitting approach based on PARSEC stellar evolutionary models was implemented to estimate 
surface gravities ($\log g$) and constrain other fundamental stellar parameters. 
The analysis combines the photometric effective temperatures and metallicities derived in Sect.~\ref{sec:calibrations} 
with absolute visual magnitudes calculated from \textit{Gaia} DR3 distances and, where available, global asteroseismic parameters.

The resulting stellar parameters are provided in a machine-readable table for 217 stars, including $T_{\rm eff}$, 
[M/H], $\log g$, stellar mass, radius, and the adopted dwarf or giant classification. One giant, KIC~757137, was excluded from 
the Bayesian analysis because its photometric indices fall outside the validity ranges of the empirical calibrations. Nevertheless, 
this star is retained in the photometric catalog and in the asteroseismic analysis. 

The surface gravities derived from the Bayesian analysis show good overall agreement with the spectroscopic values from APOGEE-2 and APOKASC-3 (Fig.~\ref{fig:comp_logg}). For the common dwarf sample, the mean difference (this work minus APOGEE-2) is $\Delta\log g=-0.008\pm0.104$ dex, indicating no appreciable systematic offset. For the giant stars, the mean difference relative to APOKASC-3 is $-0.092\pm0.174$ dex, although the median difference is only $-0.019$ dex, suggesting that a few discrepant stars contribute substantially to the larger dispersion. These results support the overall consistency of the Bayesian surface gravities with the catalog values, particularly for the dwarf sample. { We also compared our independently derived global asteroseismic parameters 
($\nu_{\rm max}$ and $\Delta\nu$), as well as the stellar masses, radii, and ages 
obtained from the Bayesian analysis, with those reported in APOKASC-3, 
finding overall good agreement.}

\begin{figure*}[!hb]
	\begin{center}
		\includegraphics[scale=0.65]
		{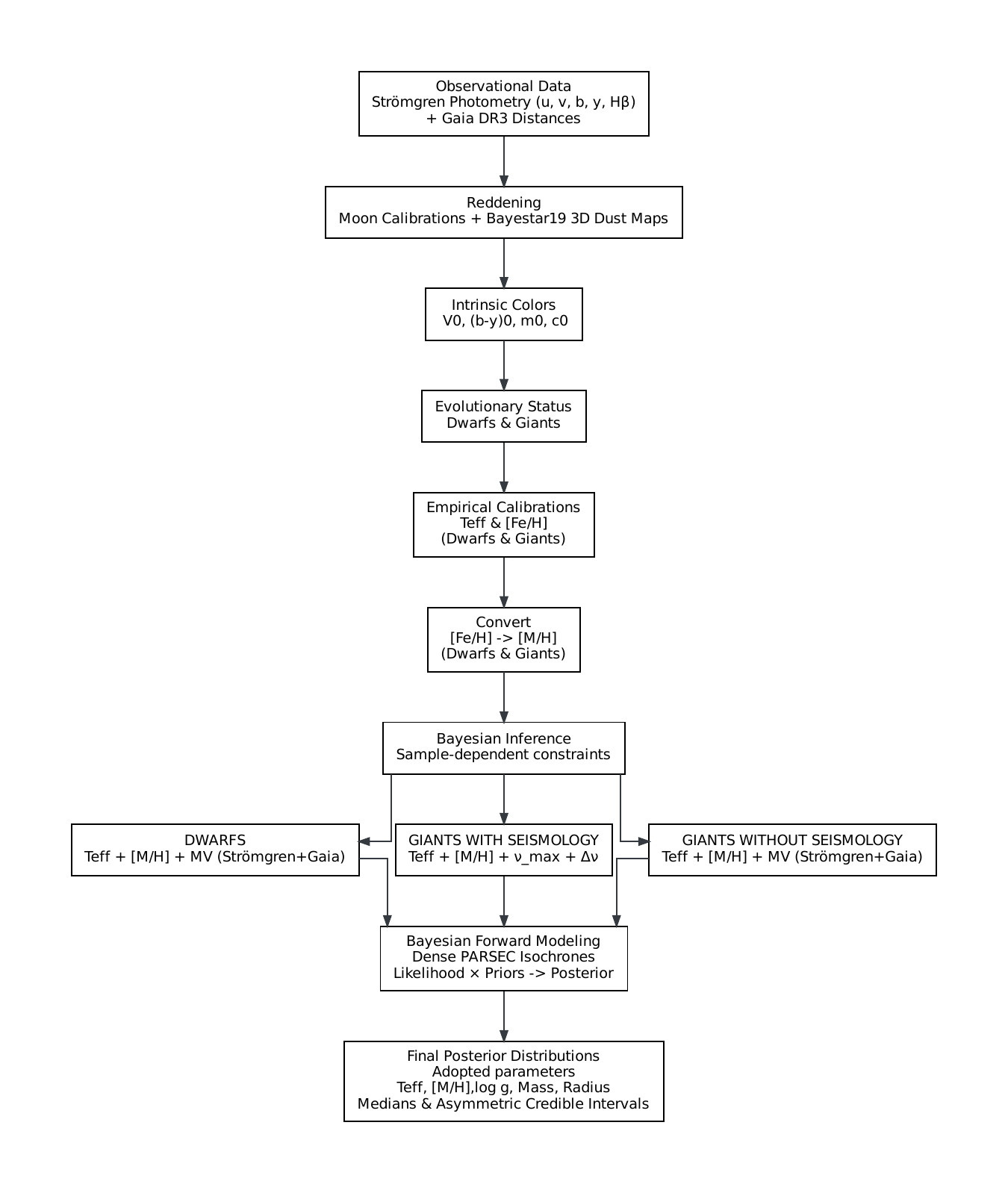}
	\end{center}
	\caption{Schematic workflow of the stellar characterization procedure adopted in this study. The diagram summarizes the main steps, from \uv photometry to the Bayesian inference of fundamental stellar parameters using PARSEC isochrones.}
	\label{fig:char}
\end{figure*}

 A schematic overview of the methodology followed in 
this study is presented in the flowchart of Fig.~\ref{fig:char}.
The methodology developed in this work will be applied to a larger sample of stars observed using CCD 
Str\"omgren photometry at the Observatorio Astronómico Nacional, San Pedro Mártir 
(OAN-SPM; Fox-Machado, L. et al., in preparation).

The Str\"omgren--Crawford photometric indices presented here provide a useful data set for further investigations of FGK-type stars in the \textit{Kepler} field, for instance, through the applications of the metaindices formalism introduced by \citet{dallemese2020}.

\begin{acknowledgments}
L.F.M. acknowledges financial support from UNAM under grant PAPIIT IN117323. This work is based on
observations carried out at the Observatorio Astronómico Nacional on the Sierra de San Pedro Mártir (OAN-SPM),
Baja California, Mexico. We thank the daytime and night support staff at OAN-SPM for
facilitating and helping obtain our observations.
This paper includes data collected by the Kepler mission and obtained from the
Mikulski Archive for Space Telescopes (MAST). Funding for the Kepler mission
was provided by the NASA Science Mission Directorate. This work has made use of data from
the European Space Agency (ESA) mission Gaia, processed by the Gaia Data Processing and Analysis
Consortium (DPAC). Funding for DPAC has been provided by national institutions,
particularly those participating in the Gaia Multilateral Agreement.
We thank the anonymous referee for valuable comments and suggestions that helped improve the quality of this paper.

\end{acknowledgments}

\begin{contribution}

The author was responsible for writing and submitting the manuscript.

\end{contribution}

%
\facilities{OANSPM:0.8m, {\it Kepler}, {\it Gaia}}


\software{
	UVBYBETA \citep{moon1985},
	Astropy \citep{2013A&A...558A..33A,2018AJ....156..123A,2022ApJ...935..167A},
	NumPy \citep{2020Natur.585..357H},
	SciPy \citep{2020NaMet..17..261V},
	Matplotlib \citep{2007CSE.....9...90H},
	Lightkurve \citep{2018ascl.soft12013L},
	Astroquery \citep{2019AJ....157...98G},
	dustmaps \citep{green2019}
}






\end{document}